\documentclass[pdftex,dvipsnames,usenames,aps,superscriptaddress,amsmath,showkeys,prd,10pt]{revtex4-2}
\usepackage{hyperref,color}

\usepackage{graphicx}
\usepackage{subfig}

\def\beq{\begin{equation}}
\def\eeq{\end{equation}}
\def\bea{\begin{eqnarray}}
\def\eea{\end{eqnarray}}
\def\figref#1{Fig.~\ref{fig:#1}}
\def\figlab#1{\label{fig:#1}}  
\def\eqref#1{Eq.~(\ref{eq:#1})}

\def\eqlab#1{\label{eq:#1}}
\newcommand*{\secref}[1]{Sec.~\ref{sec:#1}}
\newcommand*{\seclab}[1]{\label{sec:#1}}

\RequirePackage[normalem]{ulem}

\def\KVI{University of Groningen, KVI Center for Advanced Radiation Technology, Groningen, The Netherlands}
\def\Kapt{University Groningen, Kapteyn Astronomical Institute, Landleven 12, 9747 AD Groningen, The Netherlands}
\def\AIVUB{Astrophysical Institute, Vrije Universiteit Brussel, Pleinlaan 2, 1050 Brussels, Belgium}
\def\IIHE{Interuniversity Institute for High-Energy, Vrije Universiteit Brussel, Pleinlaan 2, 1050 Brussels, Belgium}

\def\IMAPP{Department of Astrophysics/IMAPP, Radboud University Nijmegen, Nijmegen, The Netherlands}

\def\ASTRON{Netherlands Institute for Radio Astronomy (ASTRON), Dwingeloo, The Netherlands}

\def\UNH{Department of Physics and Astronomy, University of New Hampshire, Durham NH 03824 USA}

\def\Erl{Erlangen Center for Astroparticle Physics, Friedrich-Alexander-Univerist\"{a}t Erlangen-N\"{u}rnberg, Germany}
\def\KIT{Institute for Astroparticle Physics, Karlsruhe Institute of Technology(KIT), P.O. Box 3640, 76021, Karlsruhe, Germany}
\def\DESY{DESY, Platanenallee 6, 15738 Zeuthen, Germany}

\def\PragIAP{Department of Space Physics, Institute of Atmospheric Physics of the Czech Academy of  Sciences, Prague, Czechia}
\def\PragCh{Faculty of Mathematics and Physics, Charles University, Prague, Czechia}

\begin{document}

\title{Interferometric imaging of Intensely Radiating Negative Leaders
}

\author{O.~Scholten}
    \affiliation{\Kapt} \affiliation{\KVI} \affiliation{\IIHE}

\author{B.~M.~Hare}
    \affiliation{\Kapt}  
\author{J.~Dwyer} %
   \affiliation{\UNH} 
\author{N.~Liu} %
   \affiliation{\UNH}   
\author{C.~Sterpka} %
   \affiliation{\UNH}  
\author{I. Kolma\v{s}ov\'{a}} %
   \affiliation{\PragIAP} \affiliation{\PragCh}  
\author{O. Santol\'{i}k} %
   \affiliation{\PragIAP} \affiliation{\PragCh}    
\author{R. L\'{a}n} %
   \affiliation{\PragIAP}     
\author{L. Uhl\'{i}\v{r}} %
   \affiliation{\PragIAP} 
\author{S.~Buitink} %
   \affiliation{\IMAPP} \affiliation{\AIVUB}  
\author{T.~Huege}     \affiliation{\KIT} \affiliation{\AIVUB} 
\author{A.~Nelles}    \affiliation{\Erl} \affiliation{\DESY}  
\author{S.~ter Veen} %
   \affiliation{\ASTRON}   


\begin{abstract}
The common phenomenon of lightning still harbors many secrets and only recently a new propagation mode was observed for negative leaders. While propagating in this `Intensely Radiating Negative Leader' (IRNL) mode a negative leader emits 100 times more very-high frequency (VHF) and broadband radiation than a more normal negative leader.
We have reported that this mode occurs soon after initiation of all lightning flashes we have mapped as well as sometimes long thereafter.
Because of the profuse emission of VHF the leader structure is very difficult to image.
In this work we report on measurements made with the LOFAR radio telescope, an instrument primarily built for radio-astronomy observations. For this reason, as part of the present work, we have refined our time resolved interferometric 3-Dimensional (TRI-D) imaging to take into account the antenna function.
The images from the TRI-D imager show that during an IRNL there is an ionization front with a diameter in excess of 500~m where strong corona bursts occur. This is very different from what is seen for a normal negative leader where the corona bursts happen at the tip, an area of typically 10~m in diameter.
The observed massive ionization wave supports the idea that this mode is indicative of a dense charge pocket.
\end{abstract}

\maketitle

\section{Introduction}

Even though lightning is very common, much of its physics is still poorly understood~\cite{Mazur:2016-ini}.
Convection inside thunderclouds separates charges into layers that result in kilometer scale strong electric fields that approach breakdown~\cite{Krehbiel:1986,Stolzenburg:2010,Trinh:2020}.
At a certain point (where it is not understood what triggers this) a discharge develops into multi-kilometer long hot conductive channels, called leaders.
It has been observed that the positively and negatively charged ends of the bi-polar discharge propagate very differently.
The negative leaders propagate in a distinct step-wise manner, emitting strong short-duration radio pulses, while positive leaders are much more radio quiet. The basic physics behind either of these modes is not understood. Once in a while such a leader may connect to ground and create a very visible return stroke.
The difficulty in exploring lightning in sufficient detail is its unpredictable nature, its violence, and the fact that it is often inside a cloud making optical observations difficult, which were, until recently, the only measurements that could reach the interesting  small scales ($<$ 1 m).

For this work we exploit radio emission in the Very-High Frequency (VHF) band to image the three dimensional (3D) lightning development in time.  VHF radiation is not obscured by the presence of clouds and is not  attenuated in the atmosphere thus allowing for large distance observations.
Radio imaging of lightning is routinely done using lightning mapping arrays \citep{Rison:1999, Edens:2012} with a temporal resolution of tens of microseconds, and has recently gone through major developments with the introduction of VHF radio interferometers \citep{Rhodes:1994, Yoshida:2010, Stock:2014}, culminating in the precision observations using the LOFAR radio telescope~\cite{Haarlem:2013} operating in the 30 -- 80~MHz band \citep{Hare:2019, Scholten:2021-init}. The main strength of LOFAR for lightning imaging lies in the combination of a large number of antennas (about 200 are customarily used out of about 3000 available), large baselines (up to 100~km), high timing stability (better than $10^{-9}$s/s), low noise level, polarization sensitivity, and the possibility for off-line analysis.

In a previous work \cite{Scholten:2021-RNL, Scholten:2021-init} we showed that negative leaders can `switch' from a normal propagation mode to an intensely radiating negative leader (IRNL) mode where they propagate faster and emit intense very-high frequency VHF (30 -- 80~MHz) and broadband radiation. This propagation mode is present only for a few milliseconds during which the VHF intensity increases by a factor 100 or 1000 and strong pulses are detected in a broadband antenna, after which normal negative leader propagation resumes, albeit that the number of negative leaders has increased considerably. The IRNL mode is seen at the initial stage of almost all lightning flashes we have imaged~\cite{Scholten:2021-init} but may also occur while the flash is propagating~\cite{Scholten:2021-RNL}.

To image the flashes in Ref.~\cite{Scholten:2021-RNL} we used our impulsive imager using data obtained from the LOFAR (Low Frequency Array) radio telescope operating in the 30 -- 80~MHz band \citep{Hare:2019, Scholten:2021-init}. The impulsive imager uses a cross-correlation procedure to determine arrival-time differences of pulses over the whole area of the LOFAR telescope and these arrival time differences are used to determine the most likely position of the source in 3D.  In spite of its meter-scale accuracy, we were not really able to distinguish the internal structure of an IRNL due to the extremely high density of emitted pulses that exceeded the confusion limit of our impulsive imager. For reference, note that this impulsive imager has no difficulty mapping many (at least up to 30) simultaneous normal negative leaders. For that reason we use for this work the recently developed  time-resolved interferometric 3-dimensional (TRI-D) imaging \cite{Scholten:2021-INL} procedure.
In TRI-D the signals from hundreds of antennas are coherently added (beamformed) for each voxel (volumetric pixel) in the imaged volume. The resulting time trace for each voxel is cut in narrow slices, where a slice may be as small as 100~ns, twice the impulse-response time of the LOFAR antennas. The coherent intensity is calculated for each time slice and each voxel. The maximum of this voxelated intensity profile is determined for each time slice and used as the source location.
With TRI-D we can resolve sources that differ in time by about the impulse-response time of our system while reaching a 3D location accuracy that is even better than the meter-scale accuracy of our impulsive imager. The main draw-back of the TRI-D imager is that it is much more computing intensive than the impulsive imager.
In addition, the TRD-D approach described in our previous work, \cite{Scholten:2021-INL},  had another draw back in that the effects of the angle dependent gain and phase-shifts due to the antenna and its electronics (antenna function) were not taken into account in the imaging procedure. This implies that antennas within a relatively narrow viewing cone can be included only in the analysis. Since some of the flashes considered in \cite{Scholten:2021-RNL} were relatively close, and thus seen under strongly varying angles by individual antennas, we have further developed the TRI-D formalism, as described in \secref{EI}, to account for the antenna function.
This has the additional benefit that future work can extend this technique to extract polarization information of the emitted radiation.

With the improved TRI-D formalism, using the data from close to 200 LOFAR antenna-pairs, we can resolve the internal structure of an IRNL. We see that IRNLs are truly completely distinct from normal negative leaders, as a typical negative leader has an almost point-like propagating tip that emits small ( $< 5$ m) corona bursts in a well-ordered fashion~\cite{Hare:2020,Scholten:2021-INL}. As shown in this work, however, an IRNL propagates by creating many corona bursts over a propagating surface of sometimes a square kilometer in size. The time trace of a broadband antenna shows a distinct sharp peak coincident with a VHF-burst recorded by LOFAR which is clear evidence for a strong sudden surge in the electric current caused by the corona burst.
An IRNL releases far more (about a factor 100)  radiation energy than during normal propagation. Because of their shear intensity, IRNLs are the prime candidates for gamma-ray production during lightning as has been observed in Refs.~\cite{Gurevich:1992, Enoto:2017, Neubert:2020, Gibney:2021, Maiorana:2021}.

\section{Methods}\seclab{Methods}

\subsection{LOFAR} \seclab{LOFAR}

LOFAR~\citep{Haarlem:2013} is a software radio telescope primarily built for radio astronomy. Signals are combined from thousands of antennas distributed over Europe in order to operate like one gigantic radio-dish.
The LOFAR antennas are arranged in stations each containing 96 inverted V-shaped dipoles in the 30-80 MHz band, referred to as Low Band Antennas (LBA) and installed in pairs with orthogonal orientations (the X- and Y- dipoles).
LOFAR also includes a similar number of High Band Antennas (100-200~MHz, that we do not use here). For data-processing reasons the data of 6 dual-polarized dipoles each from all 37 Dutch stations are recorded, although it is possible to use data from all 2000 antennas in future observations. The stations are roughly logarithmically spaced. The maximal baseline is 100~km.  
Upon an external trigger the RAM buffers are frozen that contain the raw data from every antenna. These are read out over glass fiber to  allow for off-line processing of the time traces (sampled at 200~MHz) of all antennas.

\subsection{Broadband antenna} \seclab{Bb}

For the broadband measurements we are using a magnetic loop antenna which detects the horizontal component of the time derivative of the magnetic-field vector. The Shielded Loop Antenna with a Versatile Integrated Amplifier (SLAVIA) is coupled with a digitizer sampling at 200~MHz. The frequency band of the receiver goes from 5~kHz to 90~MHz, the system has a sensitivity of 6~nT/s/$\sqrt{\rm Hz}$, corresponding to 1 fT/$\sqrt{\rm Hz}$ at 1~MHz. The setup of the receiver used in this study is in details described in \cite{Scholten:2021-RNL}. The receiver is working in a triggered mode; it records a snapshot with a duration of 167~ms whenever the absolute value of the derivative of the magnetic field exceeds a predefined threshold. The pre-trigger time is set to 52~ms.

The receiving system is installed 0.6~km to the North and 15.1~km to the East with respect to the LOFAR core, close to the village of Ter Wisch. In this installation, the antenna is most sensitive in the north-northwest direction at an azimuth of 330$^\circ$. An upward current pulse located at this azimuth with respect to the antenna produces a positive pulse in the numerically integrated magnetic field waveforms. At the Ter Wisch site, strong man-made interferences pollute substantially the recorded signal. Because of that, the waveform was cleaned using numerical narrow band rejection filters with bandwidths of 18-30~Hz at frequencies between 2 and 10~kHz, and at 18~kHz (17 filters for the 21:03 event, 19 filters for the 21:30 event).

\subsection{E-field Interferometric imaging}\seclab{EI}

The essential aspect of the updated TRI-D approach is that for each LOFAR crossed-dipole antenna the measured signals are converted to the incident polarized electric field, keeping the the complete time dependence.
Like for the original TRI-D approach the 3-D space is rasterized into voxels.
In the original TRI-D approach the interferometric intensity of a voxel was determined by coherently adding the signals of all antennas, i.e. accounting for the signal travel time. In the present approach the signals in the various antennas are calculated for a model source placed in a voxel. This model source is taken as a point-like dipole. Its strength and orientation is determined by optimizing a chi-square criterion that involves the electric field vectors at all antenna positions. This results in an expression that is very similar to the phase-shift and sum method used in the original TRI-D method.

The equations are solved in the frequency domain, i.e.\ after Fourier transforming the time-dependent signals. This makes it relatively easy to account for the antenna function. By Fourier transforming back to the time domain, the model source for each voxel is time dependent in strength and orientation. Like in the original TRI-D approach the time dependence is summed over small intervals of 100~ns, called a time slice, and for each time slice the voxel with maximal intensity is determined. The real source for this time slice is positioned at the interpolated maximum strength.

To convert the measured signals $\vec{S}_a$ on each of the dual-polarized antennas to a measured electric field the Jones matrix, $J$, is used,
\beq
\vec{E}_a = J^{-1}(\hat{r}_{as}) \vec{S}_a \;, \eqlab{EJS}
\eeq
where subscript $a$ refers to a particular antenna, $\vec{E}_a(\hat{r}_{as})$ is a 3 component vector giving the radiation electric field at the position of the antenna. $ \vec{S}_a$ is a two component vector where the two components are the measured signals in the two dipoles forming the crossed-dipole antenna (the X- and Y-dipoles) and where for  ease of notation all frequency and time dependencies are suppressed. The arrival direction of the signal is specified by $\hat{r}_{as}=\vec{r}_{as}/|\vec{r}_{as}|$ with $\vec{r}_{as}=\vec{r}_a-\vec{r}_s$ where $\vec{r}_a$ points to the antenna and likewise $\vec{r}_s$ to a particular voxel taken as the source. The unit vector  $\hat{r}_{as}$ thus points from the source to the antenna. The radiation field obeys  $\vec{E}_a \cdot \hat{r}_{as}=0$. We also introduce the distance from the source to the antenna, $R_{as}=\sqrt{\vec{r}_{as} \cdot \vec{r}_{as}}$. The Jones matrix $J$, see \secref{Jones}, parametrizes the angle and frequency dependent gain and phase-shift of the antenna and the electronics.

The radiation electric field (in the so-called far-field approximation~\cite{Jackson:1975}) at the antenna is modelled in terms of $\vec{I}$, the source current moment placed at the center of a voxel, as
\beq
\vec{E}_{as} = \frac{\vec{I} - \left(\vec{I}\cdot \hat{r}_{as}\right) \hat{r}_{as}}{R_{as}} \;,
\eeq
which obeys, by construction, $\vec{E}_{as} \cdot \hat{r}_{as}=0$ and properly falls-off with distance to the source. The fields and current moments are time dependent (but not written for simplicity of notation) and time-retardation effects are taken as implicit.

The aim of imaging is to reconstruct $\vec{I}$, including its time dependence, from the measured, time dependent, fields at the various antennas, $\vec{E}_a=\sum_{k} E_{ak} \hat{r}_{ak}$. By unfolding the antenna response the field is determined for the two, antenna dependent, polarization directions ($k={\hat{\theta}}$ and $k={\hat{\phi}}$), see \secref{Jones}. To determine the optimal value for $\vec{I}$ we formally minimize,
\beq
\chi^2_E = \sum_{a,k} \left( E_{ak} - \vec{E}_{as}\cdot \hat{r}_{ak}\right)^2 \,w_{ak}
   =\sum_{a,k} w_{ak}\left[ E_{ak}^2 - 2\, E_{ak} \left(\vec{I}\cdot \hat{r}_{ak}\right) /R_{as}  + \left(\vec{I}\cdot \hat{r}_{ak}\right)^2 /R_{as}^2   \right]  \;,
   \eqlab{chi}
\eeq
with respect to the components ${I}_i$, $i=1,2,3$. $\sum_a$ indicates a sum over all crossed-dipole antennas and $\sum_k$ implies a sum over the two orthogonal transverse polarizations $\hat{r}_{ak}$. These directions are antenna dependent where $\hat{r}_{ak}\cdot \hat{r}_{as}=0$.
Note that a proper time retardation is implicit. Antenna and polarization dependent weights $w_{ak}$ have been introduced. These weights should in principle reflect the accuracy in determining $E_{ak}$ which depends on the signal-to-noise ratio. For imaging, when searching for the location of the maximum intensity, this will complicate our algorithm too much and we have chosen to take constant weights and ignore the antenna and polarization dependence. Once the source location has been determined these weights may be implemented to determine the polarization of the source, $\vec{I}$, more accurately which is not done for this work.

Minimizing $\chi^2_E $ gives us the three conditions,
\beq
0= \partial \chi^2_E / \partial I_i = 2 \sum_{a,k} w_{ak}\left[- E_{ak} + \left(\vec{I}\cdot \hat{r}_{ak}\right)/R_{as} \right] \hat{r}_{ak,i} /R_{as}  \quad \rm{for}\; i=1,2,3 \;.
\eeq
This can be written more compactly as,
\beq
A \vec{I}= \vec{F} \;, \eqlab{AIF}
\eeq
with
\beq
\vec{F}=\sum_{a,k}  E_{ak}\,\hat{r}_{ak}\,w_{ak}/R_{as} \;, \eqlab{F}
\eeq
which is the coherent sum over all antennas (taking into account the proper time delay due to the travel time from the source to each antenna) of the fields, and
\beq
A_{ij}= \sum_{a,k} \left(\hat{r}_{ak,i} \, \hat{r}_{ak,j}\right) \,w_{ak}/R_{as}^2 \;,
\eeq
which is positive definite and symmetric. While the currents and the electric fields are time-dependent, $A_{ij}$ is not. The matrix can easily be inverted.

The current moment for a source at the center of the voxel can thus be written as
\beq
\vec{I}=A^{-1} \vec{F} \;, \eqlab{IAF}
\eeq
where the sum in $\vec{F}$ runs over all antenna positions and the proper time-shifts have to be taken into account before adding the signals from the various antennas. In spite of starting from a chi-square minimization procedure we thus have arrived at an expression that involves the coherent sum of the measured signals over all antennas. The vector $\vec{I}$ in \eqref{IAF} still contains full time and polarization information. Following again the original TRI-D approach the current is squared and summed over time slices of 100~ns to determine the interferometric intensity. Per time slice the voxel with maximal intensity (with some interpolation between nearest neighbors) is taken as image spot. In principle for each time slice the complete set of 9 Stokes parameters can be determined to retain information on the polarization of the source.

A real symmetric matrix can be written in terms of its eigenvectors, $\hat{\varepsilon}_m$, and eigenvalues, $\alpha_m$, as
\beq
A= \sum_{m=1}^3 \hat{\varepsilon}_m \alpha_m \hat{\varepsilon}_m^T \;,
\eeq
where the superscript $T$ denotes the transpose. In actual calculations it turns out that one of the eigenvalues of matrix $A$ tends to be small compared to the other two that are of similar magnitude. This results in an imbalance in the level of the noise in the three components. For this work we have therefore limited the sum over polarization to the two dominant ones. In a future work this will be investigated in more detail. Keeping the two largest eigenvalues we rewrite, for this work, \eqref{IAF} as
\beq
\vec{I}=A^{-1} \vec{F}=\sum_{m=1,2} \hat{\varepsilon}_m \alpha_m^{-1} \sum_{a,k}  \left( \hat{\varepsilon}_m \cdot \hat{r}_{ak} \right) E_{ak} \,w_a/R_{as} \;, \eqlab{IAFt}
\eeq
This corresponds to selecting the transverse component from the current moment $\vec{I}$, as seen from the core of LOFAR where the antenna density is largest.

The interferometric intensity, $I_{\rm intf}$, of each voxel is calculated for this work as
\beq
I_{\rm intf}=\sum_{\rm slice} \| \vec{I}(t) \|^2 dt\;, \eqlab{IInt}
\eeq
where $\vec{I}$ is calculated from \eqref{IAFt} taking equal weights for all antennas and the sum runs over the time samples in the time slice which is taken as 100~ns long for this work.

The weights are normalized as
\beq
w_{ak}=(R_{rs})^2/\sqrt{\sum_d \|J(\hat{r}_{rs})\|^2} \;. \eqlab{weight}
\eeq
The subscript $r$ denotes the reference antenna and $J$ is the Jones matrix as introduced in \eqref{EJS}.
The normalization of the weights in \eqref{weight} is taken such that the intensities are of the same order as those of Ref.~\cite{Scholten:2021-INL}.

\section{Data Analysis}\seclab{Analysis}

To investigate the structure of an IRNL we have made detailed interferometric images. With the interferometric imager we can even find those sources for which the pulses is not clearly separated in the time traces. Thus, even for sections of the time trace where the pulse density is extremely large, i.e.\ the trace resembles that of random noise, the interferometric imager will be able to locate the majority of the strong sources. Due to the intense background only strong sources can be located and one has to be very careful with imaging artifacts which may create spurious sources close to the edges of the image cube, caused by side beams of strong sources that lie outside the image.

In this work we show the results from the TRI-D imager for the initial stage of Flash A from \cite{Scholten:2021-RNL} in \secref{19A-5} while in \secref{19A-4} we show the internal structure imaged with TRI-D of part of the IRNL that occurred half way during the development of Flash B from  \cite{Scholten:2021-RNL}.

\subsection{Initial leader in flash A.}\seclab{19A-5}

\begin{figure*}[th]
\setlength{\unitlength}{.49\textwidth}
\begin{picture}(1,1.3)
\put(0,0){\includegraphics[bb=1.0cm 1.5cm 22.6cm 24.5cm,clip, width=0.49\textwidth]{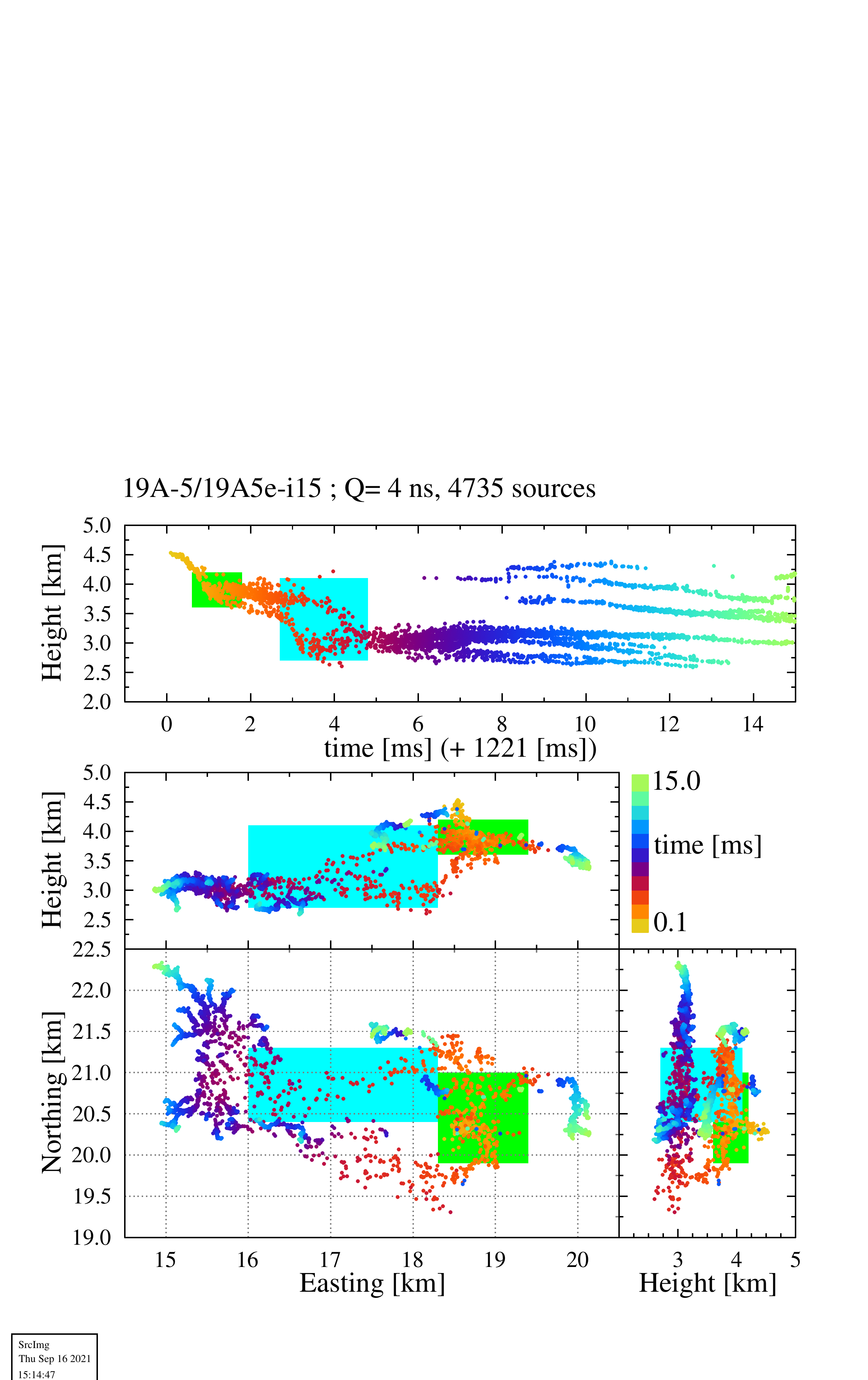}} 
\put(0,1.05){\includegraphics[bb=0.5cm 12.9cm 22.1cm 19.1cm,clip, width=0.49\textwidth]{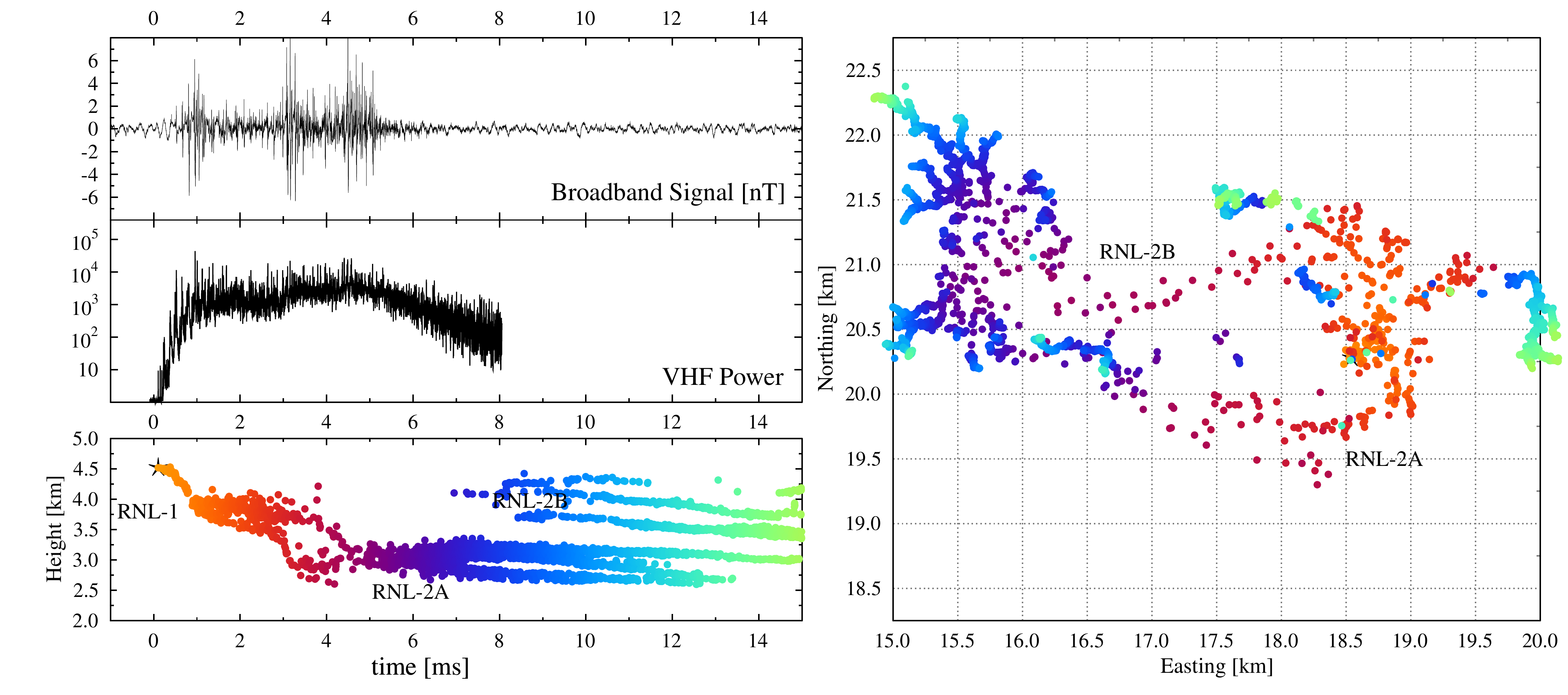}} 
\end{picture}
	\caption{Image made with the impulsive imager of the first 15 milliseconds of flash A.  The topmost panel shows the recorded signal in the broadband antenna,  re-binned over 0.2~$\mu$s. The lime-green (cyan) colored rectangles indicate the  tesseract (4 dimensional cube) imaged with TRI-D and displayed in \figref{TRID-19A5-Ini2} (\figref{TRID-19A5-Fila-N}) Respectively.  }	 \figlab{19A5-ini}
\end{figure*}

Since we have already shown the general structure of flash A in Refs.~\citep{Scholten:2021-init, Scholten:2021-RNL}, we focus here on the first 15~ms only. The time calibration is such that $t=0$ is close to the beginning of the flash at 21:30:56.221 UTC on April 24, 2019.

The height vs.\ time panel in \figref{19A5-ini} shows that immediately after initiation the initial leader propagated downward and developed into an IRNL already at t=0.5~ms when it enters the lime-green-colored tesseract imaged in \figref{TRID-19A5-Ini2} (shown as rectangles when projected on the different panels). This first IRNL developed into two separate IRNLs where the more northern one, indicated in cyan, is discussed in \secref{IRNL2}. We have selected this IRNL for imaging with TRI-D as it is still relatively compact and thus easier to image. The northern and the southern IRNLs merged around t=5~ms and a large number of normal negative leaders are seen to emerge around t=7~ms. The distinguishing features between normal negative leaders and IRNLs are~\cite{Scholten:2021-RNL} that
\begin{itemize}
\item during the IRNL mode strong VHF and broadband (topmost panel) emission is seen.
\item during IRNL propagation the impulsive imager find rather scattered sources only. This is in stark contrast with the images of normal negative leaders which are imaged as linear structures with densely packed sources.
\item during an IRNL the propagation is fast which shows in the image as sources that spread over large distances with hardly visible change in color. Normal negative leaders show a clear change in color in this visualization.
\end{itemize}
These points will be revisited in more detail when discussing later figures.

\subsubsection{The IRNL at t=0.5~ms}\seclab{IRNL1}

\begin{figure}[h]
\setlength{\unitlength}{.31\textwidth} 
	\centering	 \subfloat[IRNL, 0.6 -- 0.9 ms]{ \begin{picture}(1,1.55)
\put(0,0){\includegraphics[bb=1.0cm 1.5cm 23.1cm 30.3cm,clip, width=\unitlength]{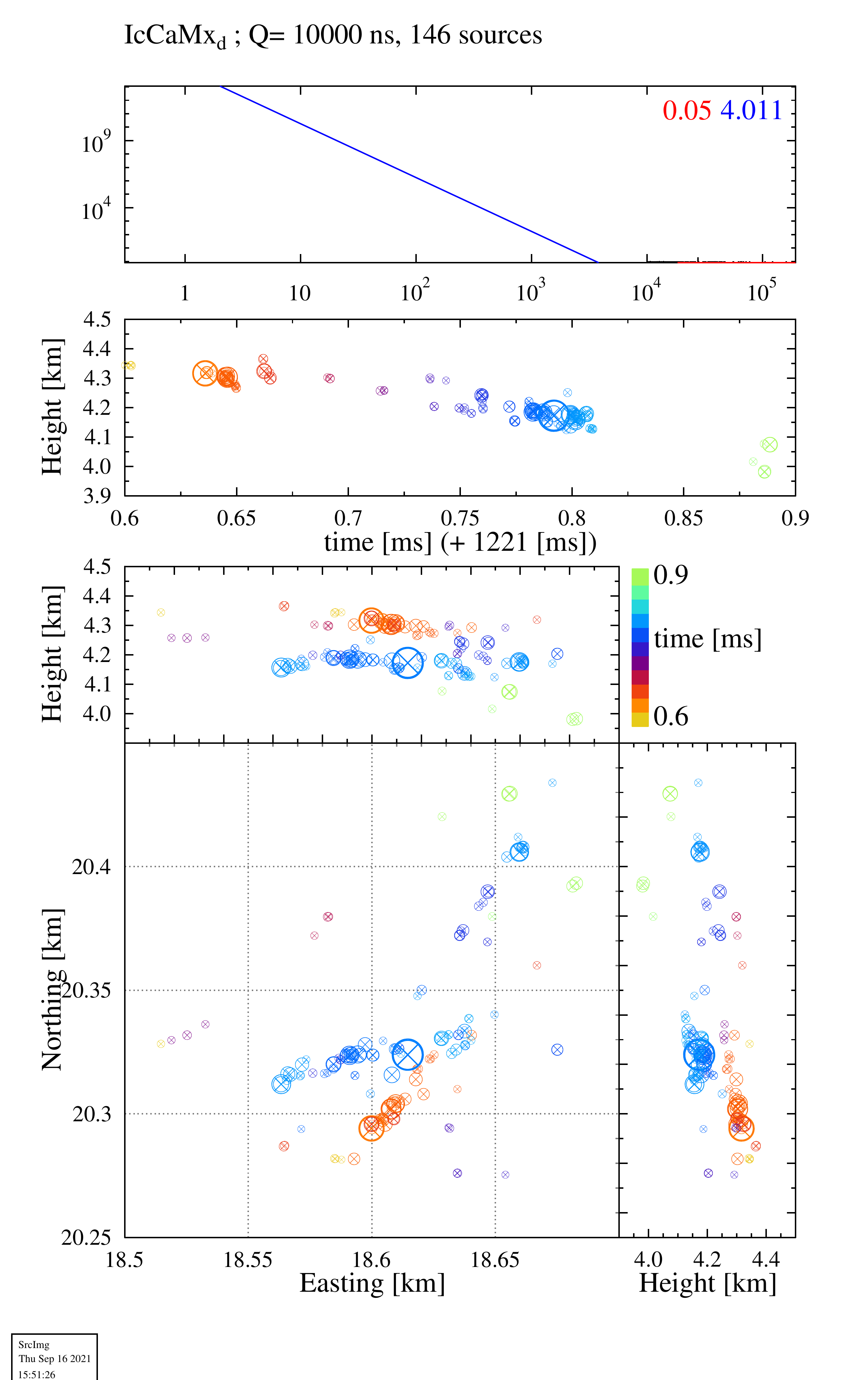}} 
\put(0,1.29){\includegraphics[bb=0.5cm 12.9cm 22.6cm 19.1cm,clip, width=\unitlength]{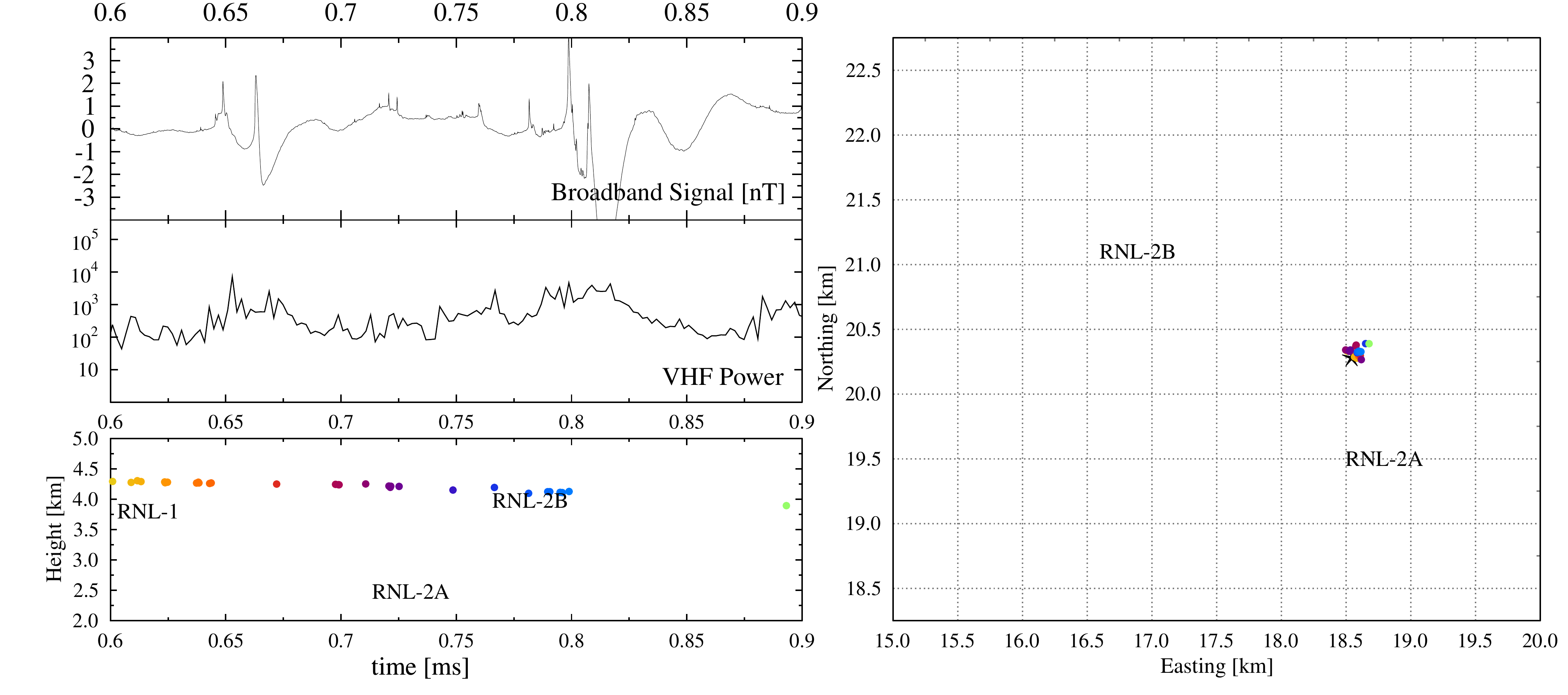}} 
\end{picture} \figlab{TRID-19A5-Ini2-a}}
\subfloat[IRNL, 1.6 -- 1.8 ms]{ \begin{picture}(1,1.55)
\put(0,0){\includegraphics[bb=1.0cm 1.5cm 23.1cm 30.3cm,clip, width=\unitlength]{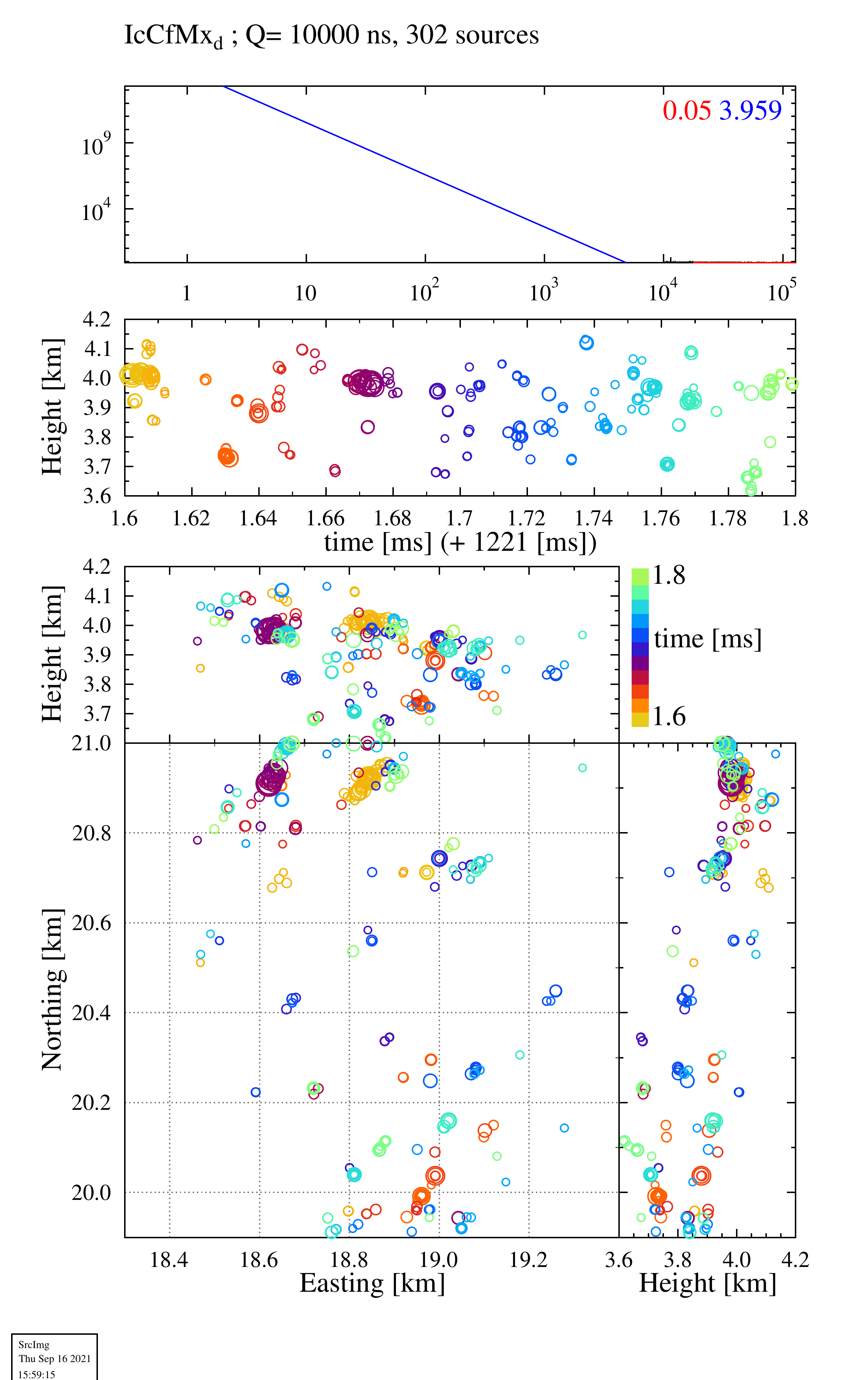}} 
\put(0,1.29){\includegraphics[bb=0.5cm 12.9cm 22.6cm 19.1cm,clip, width=\unitlength]{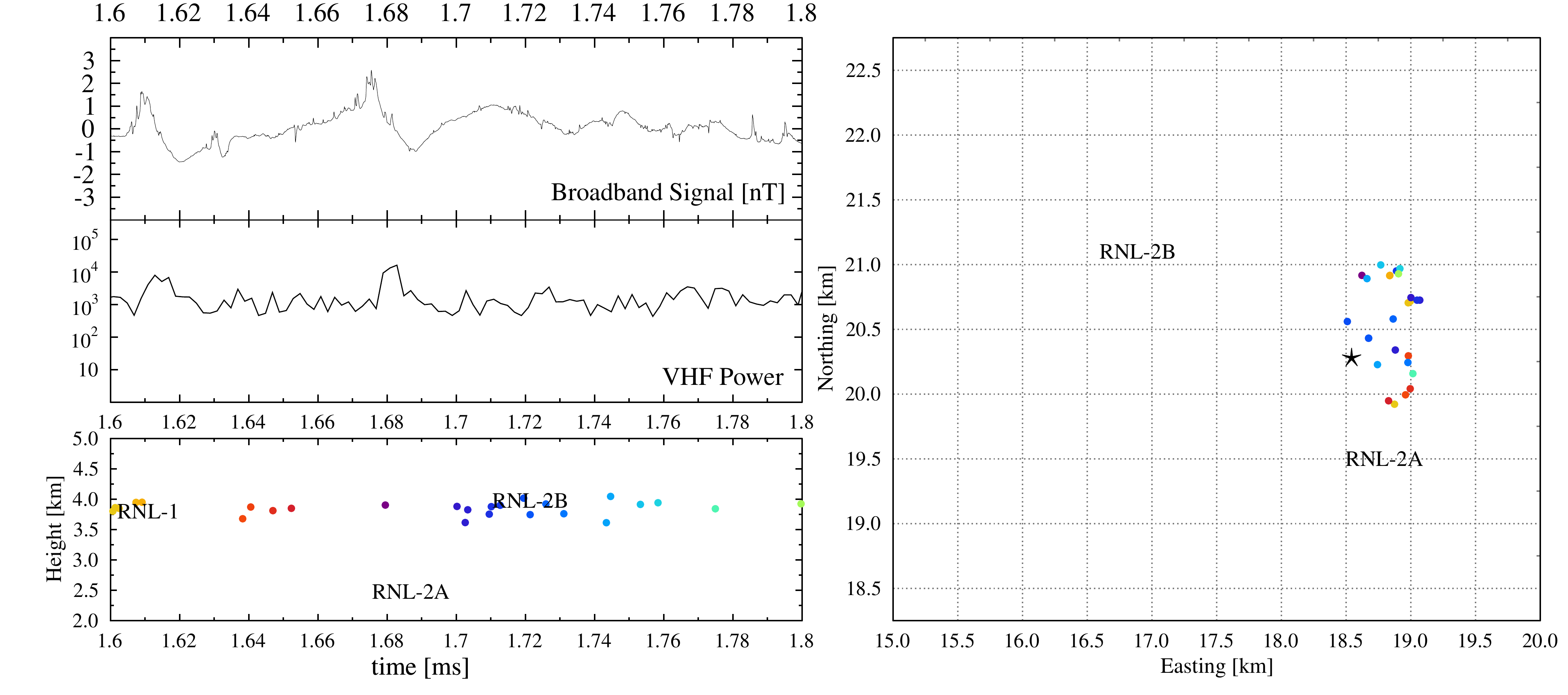}} 
\end{picture} \figlab{TRID-19A5-Ini2-e}}
\subfloat[Normal negative leader]{\includegraphics[bb=1.0cm 1.5cm 23.1cm 30.3cm,clip, width=\unitlength]{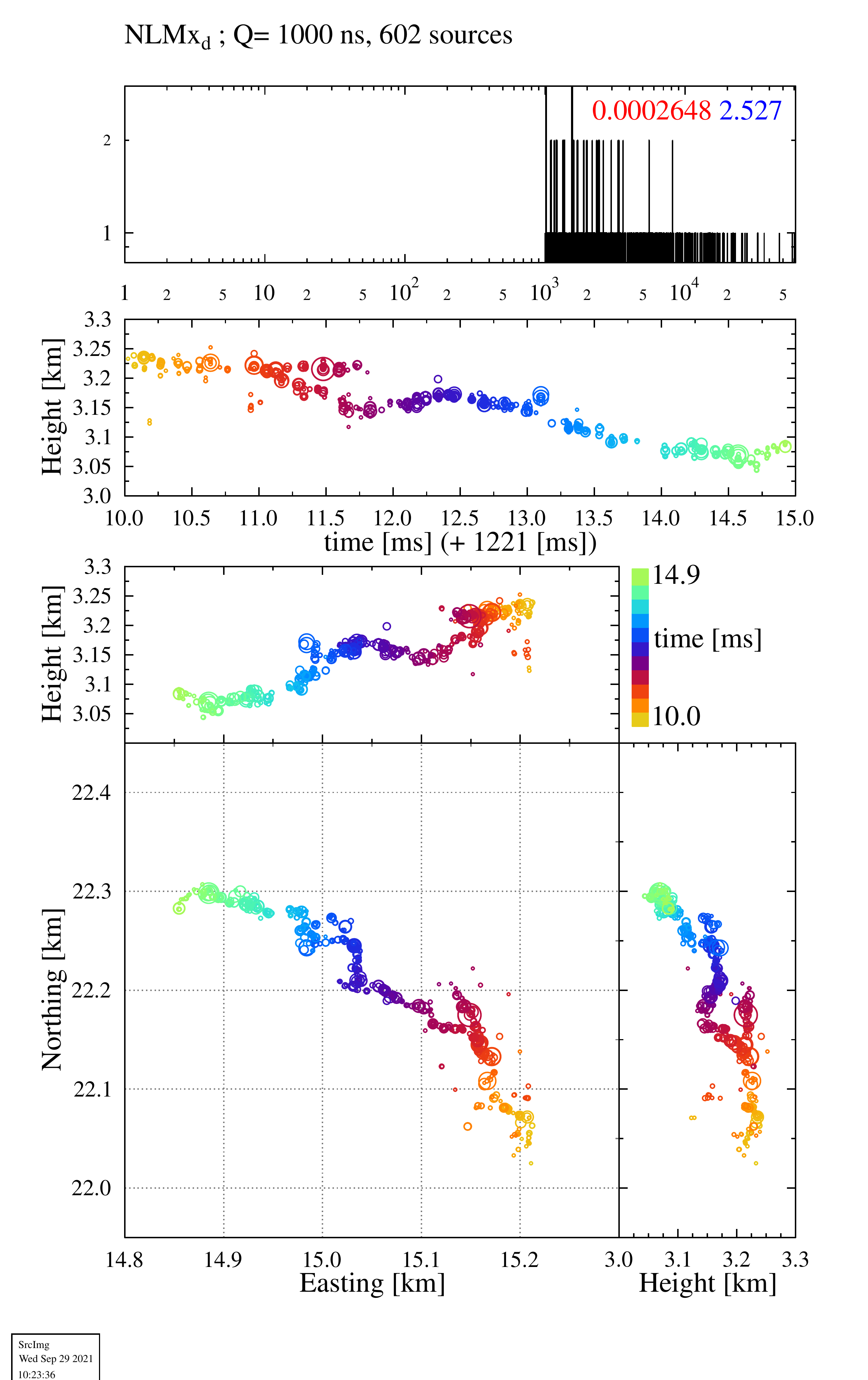} 
\figlab{TRID-19A5-Ini2-NL}}
	\caption{The left two panels show two TRI-D images of the initial part of the first IRNL seen in flash A (see  \figref{19A5-ini}, lime-green colored tesseract). For comparison the TRI-D image of a normal negative leader, propagating over a distance of about 0.5~km in 5~ms, is shown on the right. The broadband spectrum is not shown as at this time period there are several other leaders propagating.}
	\figlab{TRID-19A5-Ini2}
\end{figure}

After t=0.5~ms the initial negative leader became very bright as is evident from the broadband spectrum shown in the top panel of \figref{19A5-ini} and showed the signatures of an IRNL. Due to the very large pulse density the impulsive imager is not able to accurately image the leader structure while the TRI-D imager gives very detailed images as shown in \figref{TRID-19A5-Ini2} where we show the VHF-source locations for the early and the later part of the first IRNL in separate panels. At later times during the IRNL, the pulse intensity is of such an order that the antennas near the core are reaching saturation which makes it difficult even for the TRI-D imager. The area of the circles is proportional to the ratio of the interferometric VHF and the threshold intensity, where for the two images the same threshold intensity is used.

\figref{TRID-19A5-Ini2} clearly shows that the number of, as well as the volume filled by bright sources increases strongly during the IRNL mode. The density of bright sources stays almost constant till the end of the IRNL mode where the density starts to decrease and separate negative leaders start to form.

At early times in the IRNL the different sources can be distinguished and show as separate structures in the height vs.\ time panel. Already at this stage it is difficult to discern a clear leader structure as (almost) simultaneous sources may be separated by a few 100 meter, as seen -for example- for the light blue sources at t=0.8~ms in \figref{TRID-19A5-Ini2-a}.

The structure of the pulses seen in the broadband antenna are closely linked to the spatio-temporal development of the currents \cite{Jackson:1975,Kaspar:2015}.
The short, 1~$\mu$s, pulses seen in the time trace of the broadband antenna in \figref{TRID-19A5-Ini2-a} (such as the ones at t=0.649, 0.664, 0.798, and 0.807~ms) are clearly linked to surges in the number of strong sources seen in VHF and are likely due to corona burst. As at this stage the leader is propagating downward, a sudden increase in downward going negative charge (i.e.\ a corona burst) corresponds to an upward current producing a positive pulse for this source location with respect to the broadband antenna (to the north). As argued in \cite{Scholten:2021-INL}, a corona burst is followed by a number of streamers that gradually slow-down with limited VHF emission, even though current keeps flowing. In the broadband antenna this gives rise to a broad negative pulse. The time duration of the broad pulse, about 10~$\mu$s, is set by the slowing down time of the current through these streamers, while the negative sign is a signature of a decreasing downward current.

For distances far from the source region, the azimuthal magnetic field may be calculated from the total vertical current moment and its derivative \cite{Uman:2001}. The former is called the induction term and the latter the radiation term.  For current moments that vary on times scales of a few tens of microseconds, the radiation and induction terms will be of comparable sizes at distances on the order ten kilometers. Depending on how the current moment varies with time, the azimuthal magnetic field at the antenna can then swing between positive and negative values.  As can be seen in the upper panel of \figref{TRID-19A5-Ini2-a}, between 0.8 and 0.9 ms, the magnetic field contains several oscillations.  It is not clear if these oscillations are due to a real variation in the current moment, such as from a change in the slope of the current moment, or due to an instrumental effect where we have checked that it is not an artifact due to the cleaning of the time trace from low-frequency noise.  If it is in fact due to a real variation in the current moment, it is not obvious at this time what would be the cause.

Also for the most intense VHF sources at t=1.605~ms and at 1.67~ms in \figref{TRID-19A5-Ini2-e} there is a clear associated pulse structure seen in the Broadband antenna, similar to what was seen in \figref{TRID-19A5-Ini2-a}. However, for most of the sources this relation between VHF activity and broadband pulses is lost, probably because of the high density of pulses.

The expansion in space when the IRNL is in progress is seen most clearly by comparing \figref{TRID-19A5-Ini2-a} and \figref{TRID-19A5-Ini2-e}. While in \figref{TRID-19A5-Ini2-a} the (almost) simultaneous sources were covering an area with a diameter of 100~m, this scale has increased to 1~km in \figref{TRID-19A5-Ini2-e} where almost simultaneous strong corona bursts are seen over the whole imaged volume extending over 1~km$^2$ in the ground plane and several 100~m in height. In spite of the strong increase in the number of sources, the range of interferometric intensities, $I_{\rm intf}$, see \eqref{IInt}, is the same for the two figures.

\figref{TRID-19A5-Ini2} thus shows that an IRNL is a massive discharge. It is not possible to assign a leader structure for this IRNL simply because the density of strong sources is too high. The sources tend to be 100~m apart which is the estimated length for streamers.
At the northern fringes of the imaged area one finds spots with repeated activity which are the locations where at later times normal negative leaders will be spawned.

From \figref{TRID-19A5-Ini2-a} and \figref{TRID-19A5-Ini2-e} a clear picture emerges where the front of the IRNL shows as an increasingly large surface (or volume) over which bright VHF sources show.  This in sharp contrast with a normal negative leader shown in \figref{TRID-19A5-Ini2-NL}, where a new sources appear at the tip of a one-dimensional growing structure. Additionally the sources for a normal negative leader are less bright. The threshold intensity used for \figref{TRID-19A5-Ini2-NL} is a factor 10 lower than that for \figref{TRID-19A5-Ini2-a} and \figref{TRID-19A5-Ini2-e}.
The difference in scale between an IRNL and a normal negative leader is seen by comparing the scales of the images shown in  \figref{TRID-19A5-Ini2-e} and  \figref{TRID-19A5-Ini2-NL}. While the IRNL after a mere 1.8~ms after initiation of this flash covered an area with a diameter of about 1~km, the negative leader covered a distance of only 0.5~km after propagating for 5~ms. The diameter of the propagating head of the negative leader is smaller than 10~m.

\subsubsection{The northern part of the second IRNL}\seclab{IRNL2}

\begin{figure}[h]
\setlength{\unitlength}{.6\textwidth}
	\centering	 { \begin{picture}(1,1.2)
\put(0,0){\includegraphics[bb=1.0cm 1.5cm 23.1cm 21.8cm,clip, width=0.6\textwidth]{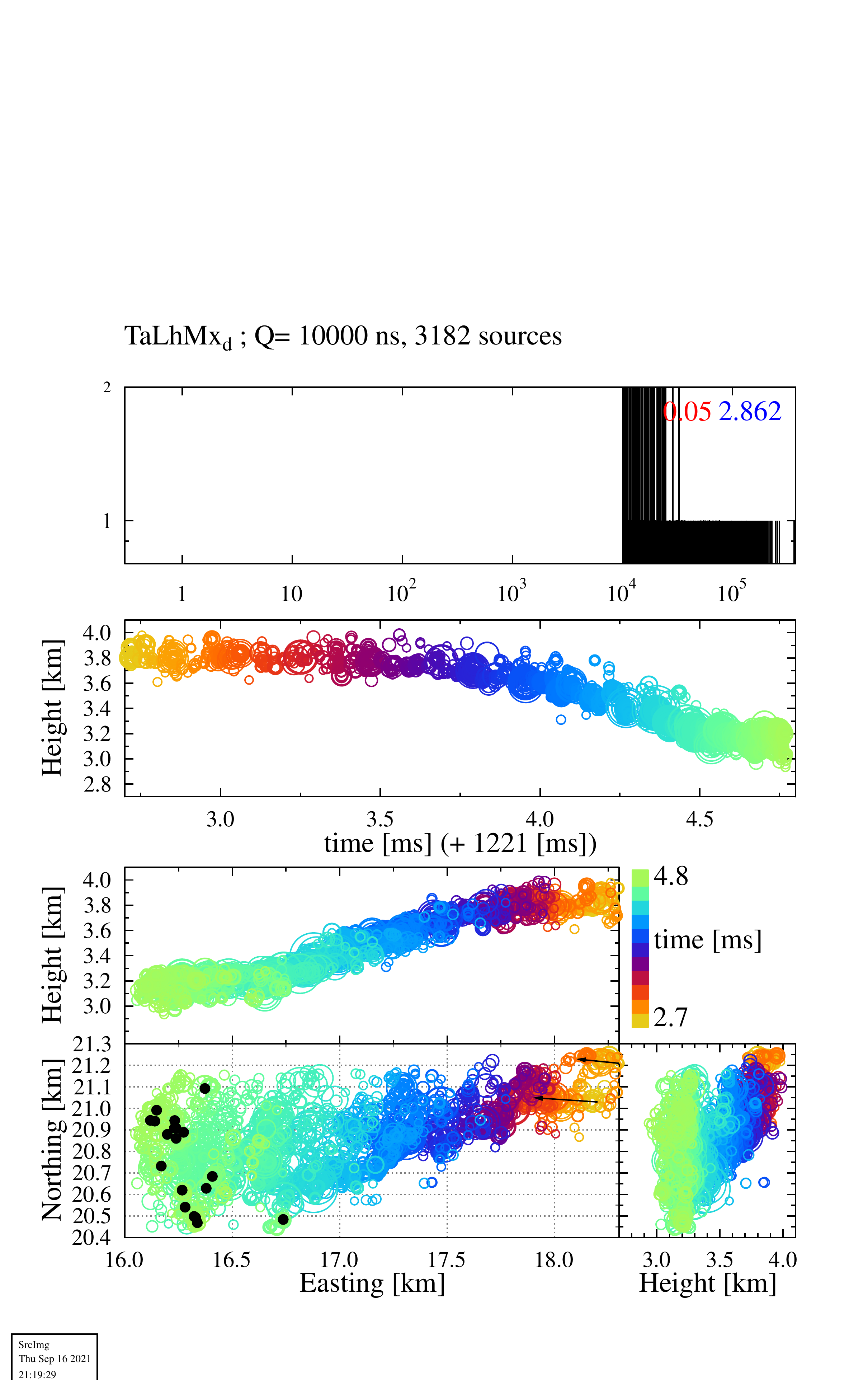}} 
\put(0,.902){\includegraphics[bb=0.5cm 12.9cm 22.6cm 19.1cm,clip, width=0.6\textwidth]{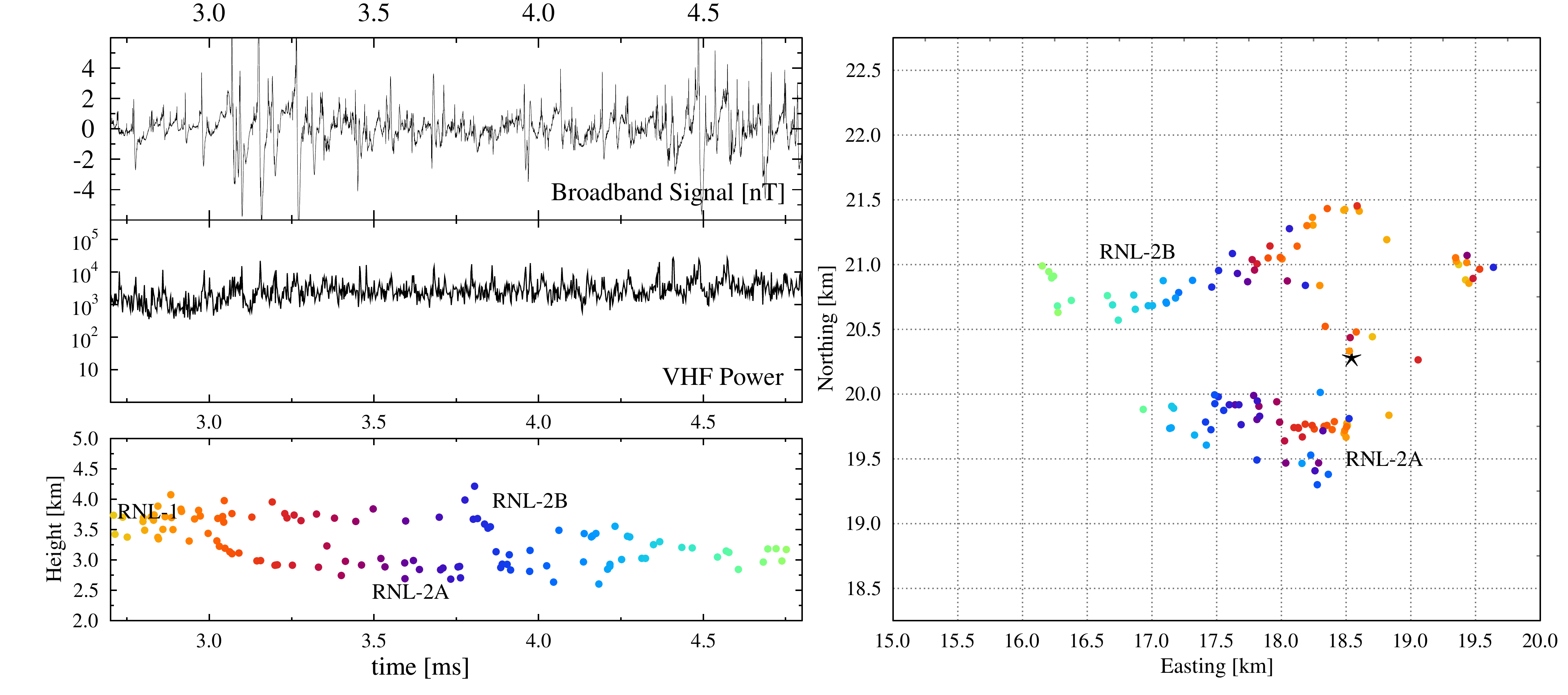}} 
\end{picture} }	
\caption{Northern filament that emerges from the first IRNL  (see  \figref{19A5-ini}). The black dots label the most recent 0.1~ms strong sources, the arrows the two starting negative leaders. }
	\figlab{TRID-19A5-Fila-N}
\end{figure}

Towards the end of this first IRNL mode two filamentary structures are seen to emerge at t=3.5~ms in \figref{19A5-ini}, that both propagate westward and down, beside some clear negative leaders. \figref{TRID-19A5-Fila-N} shows the northern filament in detail as imaged with TRI-D. Two leaders emerge from the first IRNL, marked with arrows in the figure, moving in parallel within a distance of less than 300~m. The more northern of these two bends out of the imaged area to become a normal negative leader. The main one we follow here, initially looks like a normal stepping negative leader with VHF-activity at its tip only, with occasional side branches. In the time span 2.7 -- 3.0~ms it is seen to propagate over 200~m yielding a velocity of $8 \times 10^5$~m/s which is at the upper end of velocities we observe for normal negative leaders. Soon thereafter it starts to proliferate (burgundy-maroon circles) with strong VHF emission almost simultaneously from several sources. At this point it becomes impossible to assign a clear channel structure. This structure, with many almost simultaneous sources spreading over a large surface continues. At t=4.75~ms almost simultaneous activity is observed over a volume of about (N,E,h)= (20.5 -- 21.1 , 16.2 --16.5, 3.0 --3.3)~km, as indicated by the black dots showing the source positions during the last 0.1~ms.
When comparing with the signals detected by the broadband antenna one should realize that in the initial time period a significant amount of radiation was probably emitted from the more southern leader that is not depicted in \figref{TRID-19A5-Fila-N}.

The IRNL depicted in \figref{TRID-19A5-Fila-N} covers a distance of about 2~km in 2.1~ms, giving a mean propagation velocity of $10^6$~m/s, that is an order of magnitude faster that the negative leader shown
in \figref{TRID-19A5-Ini2-NL}, propagating at $10^5$~m/s. This is yet another signature of the energy released during an IRNL.

\subsection{The IRNL of flash B at t=276~ms}\seclab{19A-4}

\begin{figure*}[th]
\setlength{\unitlength}{.6\textwidth}
\begin{picture}(1.,1.68) 
\put(0,0){\includegraphics[bb=1.0cm 1.5cm 22.6cm 31.7cm,clip, width=0.6\textwidth]{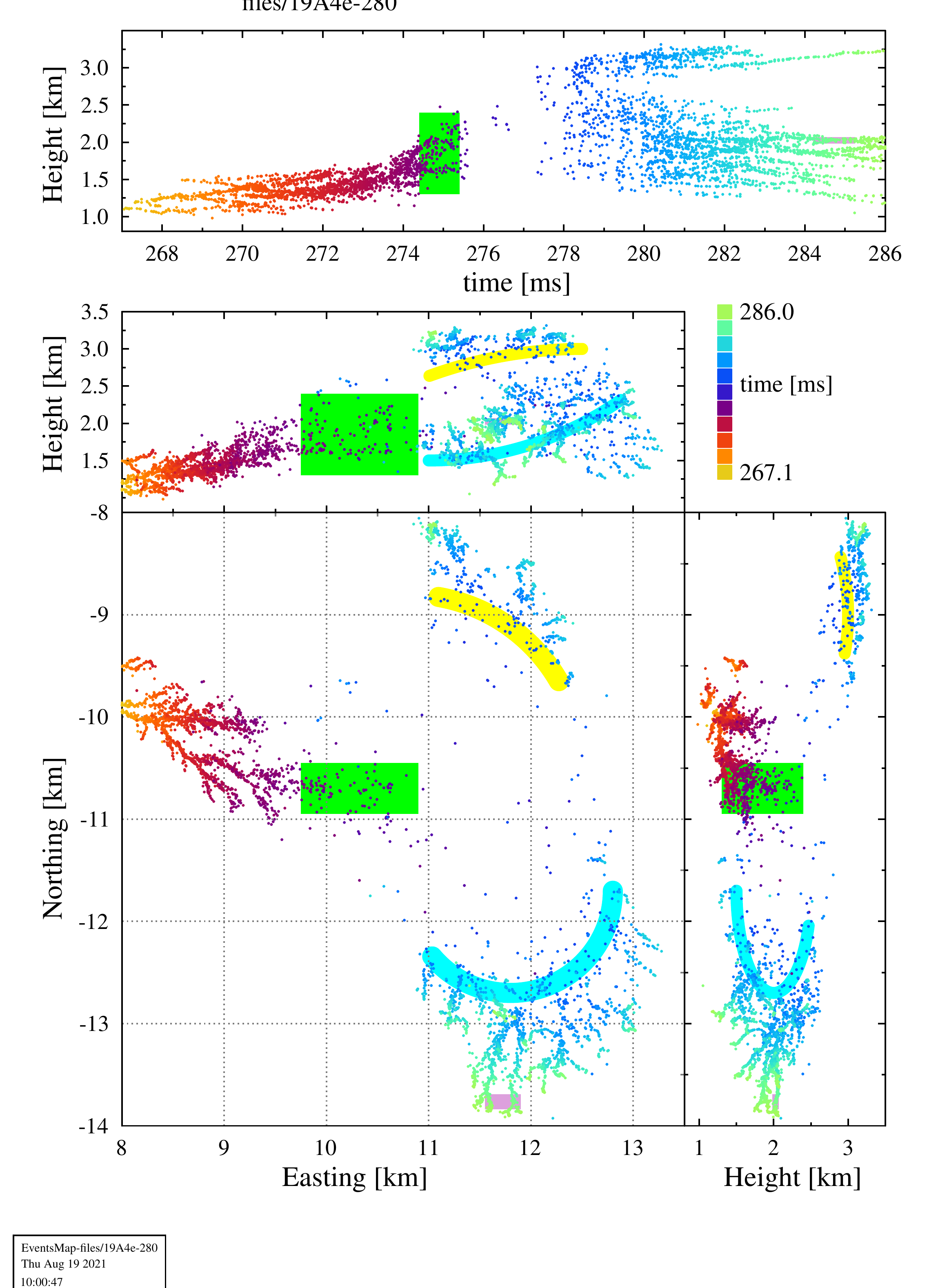} } 
\put(0,1.375){\includegraphics[bb=1.0cm 12.9cm 22.6cm 19.1cm,clip, width=0.6\textwidth]{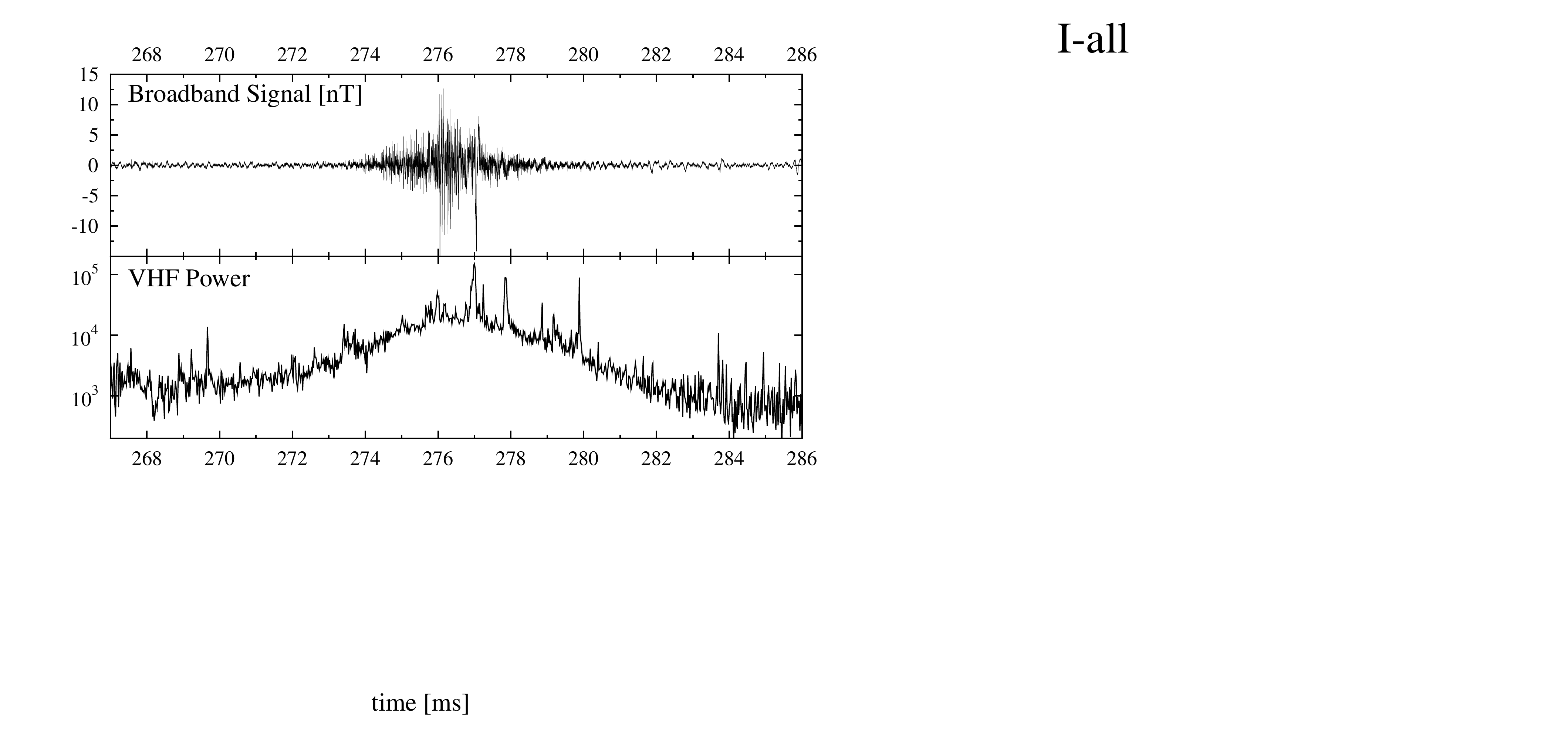} } 
\end{picture}
	\caption{An enlargement of the section around t=276~ms of flash B where strong signals were measured in the broadband antenna as shown in the top panel. The lime-green colored rectangles give the projections of the tesseract used for imaging the IRNL in \figref{TRID-19A4-R3a}, the plum colored ones that for the normal negative leader of \figref{TRID-19A4-R3b}. The yellow and cyan arcs show where the IRNL mode stops and negative leaders start to emerge.
}	\figlab{19A4-280}
\end{figure*}

For flash B, $t=0$ is chosen to be close to the beginning of the flash at 21:03:06.757 UTC on April 24, 2019. About half way during the evolution of this flash two IRNLs were observed in Ref.~\cite{Scholten:2021-RNL} some 100~ms apart at almost the same spot. In \figref{19A4-280} the first of these two re-occurring IRNLs is shown. The IRNL structure is recognizable from the large intensity seen in the broadband antenna, the disappearance of the leader structure in the image of the impulsive imager (at the lime-green colored box), the appearance of copious numbers of normal negative leaders over a large area (indicated by the yellow and cyan arcs), and the huge VHF intensity (not shown) that made it impossible for the impulsive imager to work efficiently. It should be noted that \figref{19A4-280} shows many negative leaders after the IRNL, and so demonstrates our impulsive imager's ability to locate many simultaneous negative leaders. Thus, this IRNL must be significantly more complex than even many simultaneous normal negative leaders.

\begin{figure}[h]
\subfloat[IRNL]{\includegraphics[bb=1.0cm 1.5cm 23.1cm 22.3cm,clip, width=0.5\textwidth]{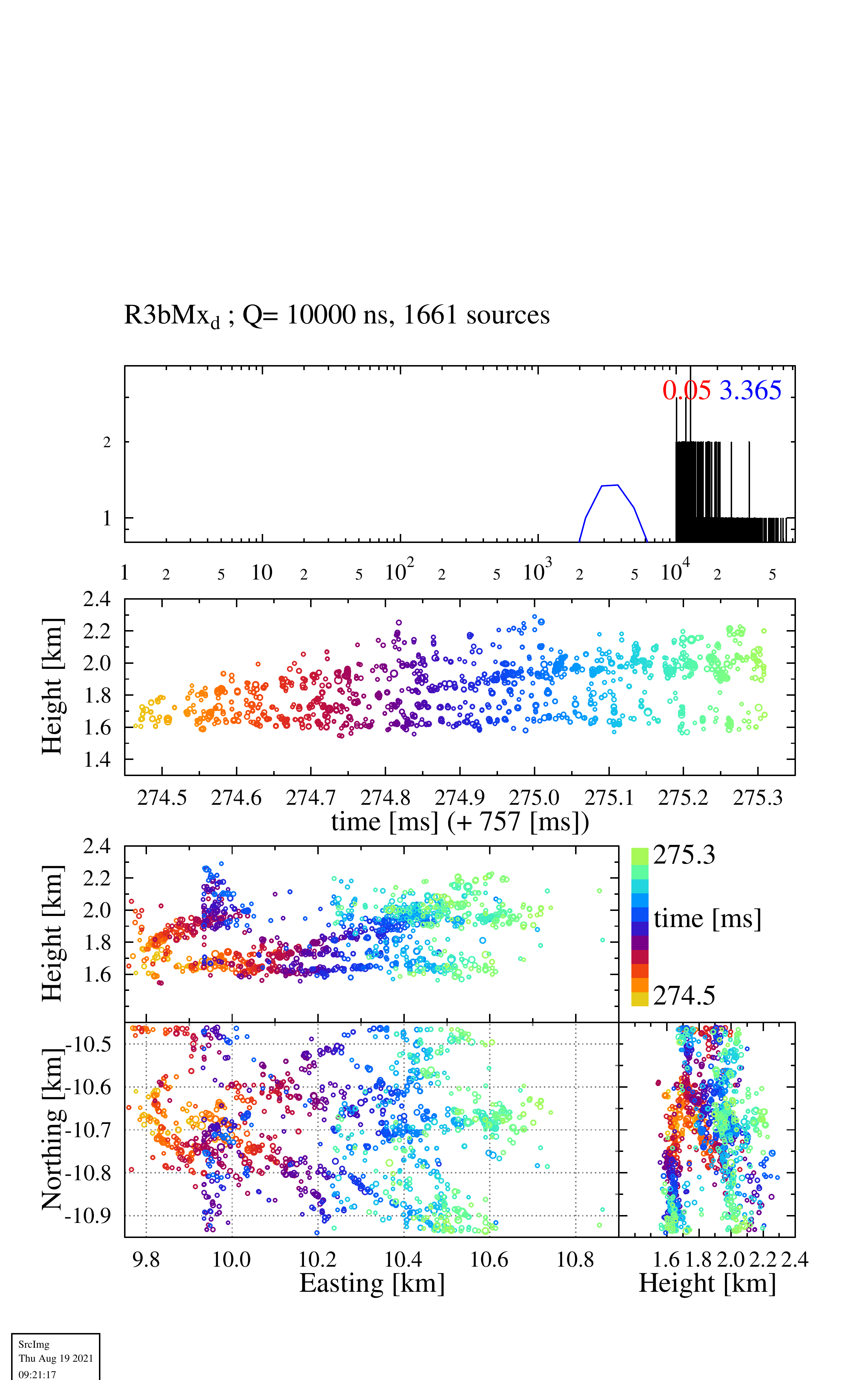} 
\figlab{TRID-19A4-R3a} }
\subfloat[normal negative leader]{\includegraphics[bb=1.0cm 1.5cm 23.1cm 30.3cm,clip, width=0.37\textwidth]{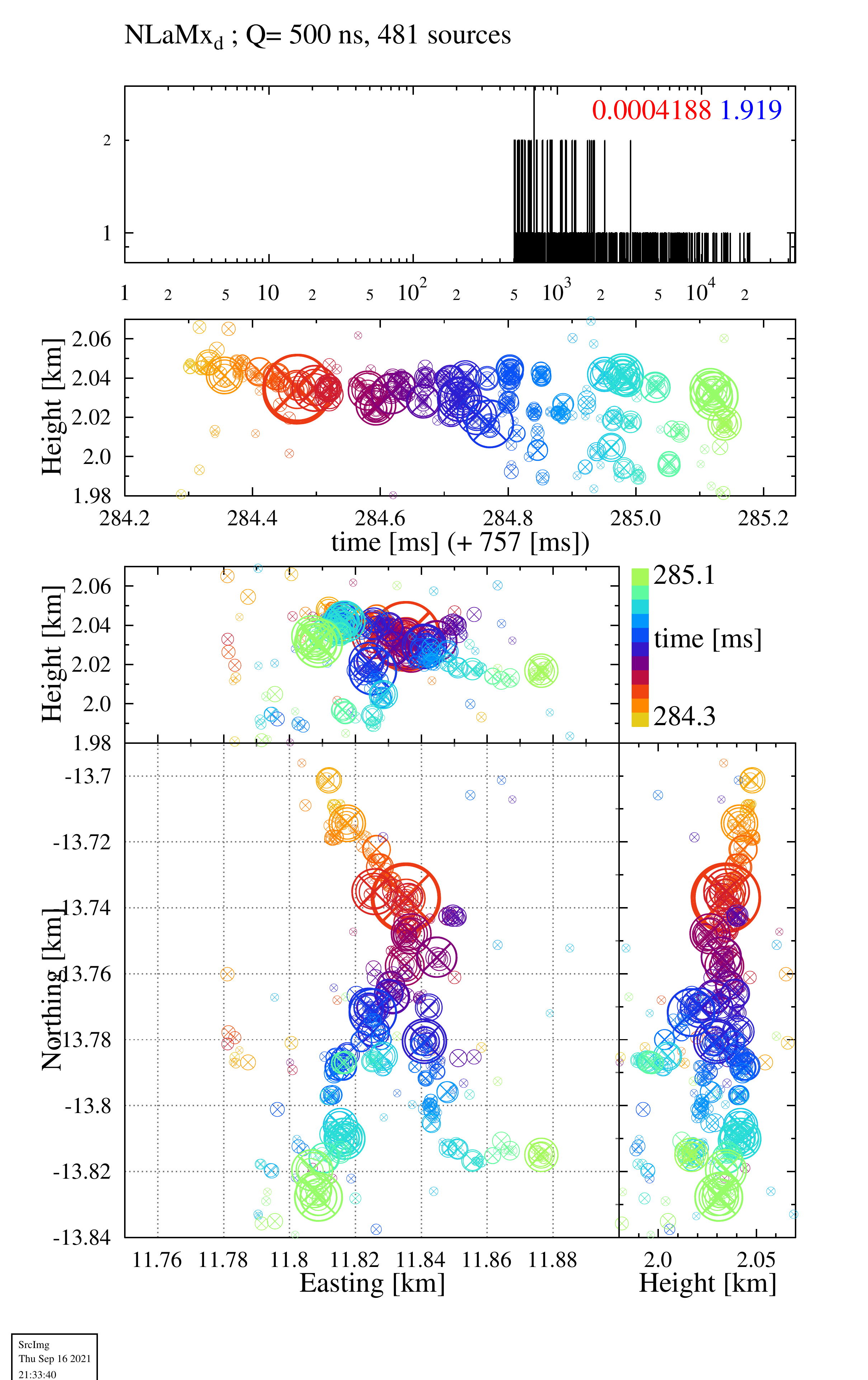} 
\figlab{TRID-19A4-R3b} }
	\caption{The TRI-D image of the early part of the IRNL for flash B is shown on the left. The imaged area is shown by the lime-green-colored tesseract in \figref{19A4-280}. The TRI-D image of a normal negative leader (indicated by the plum-colored tesseract in \figref{19A4-280}) is shown on the right to show the contrast. Note that the right-hand panels have a significantly smaller spatial scale than the left-hand panels}
	\figlab{TRID-19A4-R3}
\end{figure}

\figref{TRID-19A4-R3a} shows the TRI-D image of the IRNL shown in the lime-green-colored tesseract of \figref{19A4-280}.
In the figure shows a leader structure that at first sight still reminds of a normal negative leader, however, comparing it with the normal negative leader of \figref{TRID-19A4-R3b} clearly exemplifies the differences. They both cover the same time span during which the IRNL covers a distance of 1~km and the normal negative leader only 150~m. The IRNL seems to be constantly branching where (for practical computational reasons) some of the branches were falling outside the imaged volume. The IRNL tends to branch constantly while propagating thus creating a large number of propagating tips filling a whole surface area, rather than making a (forking) 1 dimensional leader. 

Another important difference is the intensity of the sources. The IRNL image shows 1661 sources with an interferometric intensity between $10^4$ and $7\times 10^4$, while the normal negative leader shows 481 sources with intensities between 500 and $4\times 10^4$ (27 sources with an intensity exceeding $10^4$.

\section{Summary and discussion}\seclab{Discuss}

We have updated our TRI-D imager where it now takes into account the antenna function, i.e.\ the gain and phase shifts introduced by the antenna electronics that depend on the incoming direction of the radio pulse. This allows us to alleviate the condition that the antennas have to be placed in a relatively small angular range as had to be imposed on the previous version of the TRI-D imager \cite{Scholten:2021-INL}.  Even though the formalism is written as a chi-square optimization of the received pulse (in amplitude and polarization) from a point source, the final result is expressed as a weighted coherent sum of the pulses measured in all antennas, very similar to what is obtained in interferometry. The formulation in terms of a optimization problem has the potential for generalization to more general source structures.

In a first application the updated TRI-D imager is used to image the internal structure of IRNLs, the lightning propagation mode that was only recently documented \cite{Scholten:2021-RNL}. We show that the IRNL mode is clearly distinct from a normal negative leader propagation mode. The most striking difference is that, while a negative leader has a clear propagating tip where bright sources are created in a relatively well-ordered fashion, a IRNL is propagating by creating bright sources over a rather large (often exceeding 100 $\times$ 100 m$^2$) area in copious numbers.
In one case, shown in \figref{TRID-19A4-R3a}, a large number of constantly branching negative leaders could be distinguished, filling the area of propagation, all propagating in the same direction.
Clear pulses are seen in the time trace of the broadband antenna for corona bursts during an IRNL and not for those in a normal negative leader because of the fact that the currents are much larger.
These strong currents also explain the copious emission of VHF radiation.
Additionally we observe that the propagation speed is about one order of magnitude faster than for normal negative leaders.

This finding supports the originally proposed picture where an IRNL should be regarded as a massive discharge wave passing through a pocket with high charge density. The release of orders of magnitude more energy during the IRNL mode as compared to that for a normal leader make IRNLs the prime suspects for the site of the observed gamma-ray emission during thunderstorms~\cite{Gurevich:1992, Enoto:2017, Neubert:2020, Gibney:2021}. Interesting in this respect is that in Ref.~\cite{Wada:2019} short duration pulses, seemingly similar to what we observe with the broadband antenna, were seen in coincidence with gamma ray flashes observed at ground level.

\section{acknowledgements}

BMH is supported by the NWO [grant number VI.VENI.192.071];
The work of the IAP team was supported by European Regional Development Fund-Project CRREAT [grant number CZ.02.1.01/0.0/0.0/15-003/0000481] and by the GACR [grant number 20-09671S].
\\
This paper is based on data obtained with the International LOFAR Telescope (ILT). LOFAR~\citep{Haarlem:2013} is the Low Frequency Array designed and constructed by ASTRON. It has observing, data processing, and data storage facilities in several countries, that are owned by various parties (each with their own funding sources), and that are collectively operated by the ILT foundation under a joint scientific policy. The ILT resources have benefitted from the following recent major funding sources: CNRS-INSU, Observatoire de Paris and Universit\'{e} d'Orl\'{e}ans, France; BMBF, MIWF-NRW, MPG, Germany; Science Foundation Ireland (SFI), Department of Business, Enterprise and Innovation (DBEI), Ireland; NWO, The Netherlands; The Science and Technology Facilities Council, UK.

The data are available from the LOFAR Long Term Archive (for access see \citep{LOFAR-LTA}).
To download this data, please create an account and follow the instructions for ``Staging Transient Buffer Board data'' at \citep{LOFAR-LTA}. In particular, the utility ``wget'' should be used as follows:
\\ {\footnotesize \verb!wget https://lofar-download.grid.surfsara.nl/lofigrid/SRMFifoGet.py?surl="location"!}
where ``location'' should be specified as:
\\ {\small \verb!srm://srm.grid.sara.nl/pnfs/grid.sara.nl/data/lofar/ops/TBB/lightning/!} followed by
\\ {\small \verb!L703974_D20190424T210306.154Z_"stat"_R000_tbb.h5!} (for Flash B)
\\ {\small \verb!L703974_D20190424T213055.202Z_"stat"_R000_tbb.h5!} (for Flash A)
and where
``stat'' should be replaced by the name of the station, CS001, CS002, CS003, CS004, CS005, CS006, CS007, CS011, CS013, CS017, CS021, CS024, CS026, CS028, CS030, CS031, CS032, CS101, CS103, RS106, CS201, RS205, RS208, RS210, CS301, CS302, RS305, RS306, RS307, RS310, CS401, RS406, RS407, RS409, CS501, RS503, RS508, or RS509.
The source code used for improved TRI-D imaging can be found at \citep{Scholten-v20:2021}.

All figures in this work have been made using the Graphics Layout Engine (GLE) \cite{GLE} plotting package.

\appendix

\section{The Antenna Model}\seclab{Jones}

The response of the antennas was found by simulating the LOFAR antennas in WIPL-D, including the effect of a realistic ground, interactions of nearby antennas, and coupling to the LOFAR low noise amplifier (LNA)~\cite{Arts:2021}. The results of this simulation are represented as Jones matrix which converts incident transverse oriented electric fields ($E_k$, $k=\hat{\theta}=$zenithal, $k=\hat{\phi}=$azimuthal) into a measured voltage, $\vec{S}_d$, for each of the two crossed bent $d=(X,Y)$-dipole antennas, as a function of frequency and incident angle. As such $J$ can be regarded as a $(2\times2)$ matrix,
\beq
\vec{S}_d = \sum_k J(\nu;\hat{r}_s)_{d,k} E_k\, \hat{r}_{k} \;,
\eeq
that depends on frequency, $\nu$, as well as on the arrival direction $\hat{r}_s$ of the radiation that is assumed to be a plane wave.

The difficulty is that this Jones matrix has only been calculated on a grid~\cite{Arts:2021} whose points do not necessarily correspond with the frequency and angle we need in our imaging. Therefore, we fit the simulated Jones matrices with a smooth function of the direction $\hat{r}=(\theta,\phi)$ and write it as a sum over spherical harmonics,
\beq
J_{d,k}(\nu;\theta,\phi)=\sum_{j,m} A_{j,m}^{(d,k)}(\nu)\, T_m(\sin{\phi})\, P_j(\cos{\theta})
\eeq
where $d=(X,Y)$ denotes the dipole, $k=(\hat{\theta},\hat{\phi})$ the polarization of the electric field, and $A_{j,m}^{(d,k)}(\nu)$ the functions that parameterize the Jones matrix. for the parametrization used in this work the sum runs over $j=(1,3,5,7)$ for the Legendre polynomials, $P_j$ that depend on zenith angle, and $m=(1,3,5)$ for the Chebyshev polynomials, $T_m$ that depend on azimuth angle. The X- and Y- dipoles are assumed to be identical, only rotated over 90$^\circ$.
The functions $A_{j,m}^{(d,k)}(\nu)$ are determined by fitting tabulated values of the Jones matrix \cite{Arts:2021}.

\renewcommand{\bibsection}{\subsection*{References}}
\bibliography{}

\begin{thebibliography}{28}%
\makeatletter
\providecommand \@ifxundefined [1]{%
 \@ifx{#1\undefined}
}%
\providecommand \@ifnum [1]{%
 \ifnum #1\expandafter \@firstoftwo
 \else \expandafter \@secondoftwo
 \fi
}%
\providecommand \@ifx [1]{%
 \ifx #1\expandafter \@firstoftwo
 \else \expandafter \@secondoftwo
 \fi
}%
\providecommand \natexlab [1]{#1}%
\providecommand \enquote  [1]{``#1''}%
\providecommand \bibnamefont  [1]{#1}%
\providecommand \bibfnamefont [1]{#1}%
\providecommand \citenamefont [1]{#1}%
\providecommand \href@noop [0]{\@secondoftwo}%
\providecommand \href [0]{\begingroup \@sanitize@url \@href}%
\providecommand \@href[1]{\@@startlink{#1}\@@href}%
\providecommand \@@href[1]{\endgroup#1\@@endlink}%
\providecommand \@sanitize@url [0]{\catcode `\\12\catcode `\$12\catcode
  `\&12\catcode `\#12\catcode `\^12\catcode `\_12\catcode `\%12\relax}%
\providecommand \@@startlink[1]{}%
\providecommand \@@endlink[0]{}%
\providecommand \url  [0]{\begingroup\@sanitize@url \@url }%
\providecommand \@url [1]{\endgroup\@href {#1}{\urlprefix }}%
\providecommand \urlprefix  [0]{URL }%
\providecommand \Eprint [0]{\href }%
\providecommand \doibase [0]{https://doi.org/}%
\providecommand \selectlanguage [0]{\@gobble}%
\providecommand \bibinfo  [0]{\@secondoftwo}%
\providecommand \bibfield  [0]{\@secondoftwo}%
\providecommand \translation [1]{[#1]}%
\providecommand \BibitemOpen [0]{}%
\providecommand \bibitemStop [0]{}%
\providecommand \bibitemNoStop [0]{.\EOS\space}%
\providecommand \EOS [0]{\spacefactor3000\relax}%
\providecommand \BibitemShut  [1]{\csname bibitem#1\endcsname}%
\let\auto@bib@innerbib\@empty
\bibitem [{\citenamefont {Mazur}(2016)}]{Mazur:2016-ini}%
  \BibitemOpen
  \bibfield  {author} {\bibinfo {author} {\bibfnamefont {V.}~\bibnamefont
  {Mazur}},\ }in\ \href {https://doi.org/10.1088/978-0-7503-1152-6ch12} {\emph
  {\bibinfo {booktitle} {Principles of Lightning Physics}}},\ \bibinfo {series
  and number} {2053-2563}\ (\bibinfo  {publisher} {IOP Publishing},\ \bibinfo
  {year} {2016})\ pp.\ \bibinfo {pages} {12--1 to 12--10}\BibitemShut {NoStop}%
\bibitem [{\citenamefont {Krehbiel}(1986)}]{Krehbiel:1986}%
  \BibitemOpen
  \bibfield  {author} {\bibinfo {author} {\bibfnamefont {P.}~\bibnamefont
  {Krehbiel}},\ }in\ \href {https://doi.org/10.17226/898} {\emph {\bibinfo
  {booktitle} {{The Earth's electrical environment}}}},\ \bibinfo {editor}
  {edited by\ \bibinfo {editor} {\bibfnamefont {N.~R.}\ \bibnamefont
  {Council}}}\ (\bibinfo  {publisher} {The National Academies Press},\ \bibinfo
  {address} {Washington, DC},\ \bibinfo {year} {1986})\BibitemShut {NoStop}%
\bibitem [{\citenamefont {Stolzenburg}\ \emph {et~al.}(2010)\citenamefont
  {Stolzenburg}, \citenamefont {Marshall},\ and\ \citenamefont
  {Krehbiel}}]{Stolzenburg:2010}%
  \BibitemOpen
  \bibfield  {author} {\bibinfo {author} {\bibfnamefont {M.}~\bibnamefont
  {Stolzenburg}}, \bibinfo {author} {\bibfnamefont {T.~C.}\ \bibnamefont
  {Marshall}},\ and\ \bibinfo {author} {\bibfnamefont {P.~R.}\ \bibnamefont
  {Krehbiel}},\ }\href {https://doi.org/10.1029/2010JD014057} {\bibfield
  {journal} {\bibinfo  {journal} {Journal of Geophysical Research:
  Atmospheres}\ }\textbf {\bibinfo {volume} {115}},\ \bibinfo {pages} {202}
  (\bibinfo {year} {2010})}\BibitemShut {NoStop}%
\bibitem [{\citenamefont {Trinh}\ \emph {et~al.}(2020)\citenamefont {Trinh},
  \citenamefont {Scholten}, \citenamefont {Buitink}, \citenamefont {Ebert},
  \citenamefont {Hare}, \citenamefont {Krehbiel}, \citenamefont {Leijnse},
  \citenamefont {Bonardi}, \citenamefont {Corstanje}, \citenamefont {Falcke},
  \citenamefont {Huege}, \citenamefont {H\"{o}randel}, \citenamefont {Krampah},
  \citenamefont {Mitra}, \citenamefont {Mulrey}, \citenamefont {Nelles},
  \citenamefont {Pandya}, \citenamefont {Rachen}, \citenamefont {Rossetto},
  \citenamefont {Rutjes}, \citenamefont {ter Veen},\ and\ \citenamefont
  {Winchen}}]{Trinh:2020}%
  \BibitemOpen
  \bibfield  {author} {\bibinfo {author} {\bibfnamefont {T.~N.~G.}\
  \bibnamefont {Trinh}}, \bibinfo {author} {\bibfnamefont {O.}~\bibnamefont
  {Scholten}}, \bibinfo {author} {\bibfnamefont {S.}~\bibnamefont {Buitink}},
  \bibinfo {author} {\bibfnamefont {U.}~\bibnamefont {Ebert}}, \bibinfo
  {author} {\bibfnamefont {B.~M.}\ \bibnamefont {Hare}}, \bibinfo {author}
  {\bibfnamefont {P.~R.}\ \bibnamefont {Krehbiel}}, \bibinfo {author}
  {\bibfnamefont {H.}~\bibnamefont {Leijnse}}, \bibinfo {author} {\bibfnamefont
  {A.}~\bibnamefont {Bonardi}}, \bibinfo {author} {\bibfnamefont
  {A.}~\bibnamefont {Corstanje}}, \bibinfo {author} {\bibfnamefont
  {H.}~\bibnamefont {Falcke}}, \bibinfo {author} {\bibfnamefont
  {T.}~\bibnamefont {Huege}}, \bibinfo {author} {\bibfnamefont {J.~R.}\
  \bibnamefont {H\"{o}randel}}, \bibinfo {author} {\bibfnamefont {G.~K.}\
  \bibnamefont {Krampah}}, \bibinfo {author} {\bibfnamefont {P.}~\bibnamefont
  {Mitra}}, \bibinfo {author} {\bibfnamefont {K.}~\bibnamefont {Mulrey}},
  \bibinfo {author} {\bibfnamefont {A.}~\bibnamefont {Nelles}}, \bibinfo
  {author} {\bibfnamefont {H.}~\bibnamefont {Pandya}}, \bibinfo {author}
  {\bibfnamefont {J.~P.}\ \bibnamefont {Rachen}}, \bibinfo {author}
  {\bibfnamefont {L.}~\bibnamefont {Rossetto}}, \bibinfo {author}
  {\bibfnamefont {C.}~\bibnamefont {Rutjes}}, \bibinfo {author} {\bibfnamefont
  {S.}~\bibnamefont {ter Veen}},\ and\ \bibinfo {author} {\bibfnamefont
  {T.}~\bibnamefont {Winchen}},\ }\href {https://doi.org/10.1029/2019JD031433}
  {\bibfield  {journal} {\bibinfo  {journal} {Journal of Geophysical Research:
  Atmospheres}\ }\textbf {\bibinfo {volume} {125}},\ \bibinfo {pages}
  {e2019JD031433} (\bibinfo {year} {2020})},\ \bibinfo {note} {e2019JD031433
  10.1029/2019JD031433}\BibitemShut {NoStop}%
\bibitem [{\citenamefont {Rison}\ \emph {et~al.}(1999)\citenamefont {Rison},
  \citenamefont {Thomas}, \citenamefont {Krehbiel}, \citenamefont {Hamlin},\
  and\ \citenamefont {Harlin}}]{Rison:1999}%
  \BibitemOpen
  \bibfield  {author} {\bibinfo {author} {\bibfnamefont {W.}~\bibnamefont
  {Rison}}, \bibinfo {author} {\bibfnamefont {R.~J.}\ \bibnamefont {Thomas}},
  \bibinfo {author} {\bibfnamefont {P.~R.}\ \bibnamefont {Krehbiel}}, \bibinfo
  {author} {\bibfnamefont {T.}~\bibnamefont {Hamlin}},\ and\ \bibinfo {author}
  {\bibfnamefont {J.}~\bibnamefont {Harlin}},\ }\href
  {https://doi.org/10.1029/1999GL010856} {\bibfield  {journal} {\bibinfo
  {journal} {Geophysical Research Letters}\ }\textbf {\bibinfo {volume} {26}},\
  \bibinfo {pages} {3573} (\bibinfo {year} {1999})}\BibitemShut {NoStop}%
\bibitem [{\citenamefont {Edens}\ \emph {et~al.}(2012)\citenamefont {Edens},
  \citenamefont {Eack}, \citenamefont {Eastvedt}, \citenamefont {Trueblood},
  \citenamefont {Winn}, \citenamefont {Krehbiel}, \citenamefont {Aulich},
  \citenamefont {Hunyady}, \citenamefont {Murray}, \citenamefont {Rison},
  \citenamefont {Behnke},\ and\ \citenamefont {Thomas}}]{Edens:2012}%
  \BibitemOpen
  \bibfield  {author} {\bibinfo {author} {\bibfnamefont {H.~E.}\ \bibnamefont
  {Edens}}, \bibinfo {author} {\bibfnamefont {K.~B.}\ \bibnamefont {Eack}},
  \bibinfo {author} {\bibfnamefont {E.~M.}\ \bibnamefont {Eastvedt}}, \bibinfo
  {author} {\bibfnamefont {J.~J.}\ \bibnamefont {Trueblood}}, \bibinfo {author}
  {\bibfnamefont {W.~P.}\ \bibnamefont {Winn}}, \bibinfo {author}
  {\bibfnamefont {P.~R.}\ \bibnamefont {Krehbiel}}, \bibinfo {author}
  {\bibfnamefont {G.~D.}\ \bibnamefont {Aulich}}, \bibinfo {author}
  {\bibfnamefont {S.~J.}\ \bibnamefont {Hunyady}}, \bibinfo {author}
  {\bibfnamefont {W.~C.}\ \bibnamefont {Murray}}, \bibinfo {author}
  {\bibfnamefont {W.}~\bibnamefont {Rison}}, \bibinfo {author} {\bibfnamefont
  {S.~A.}\ \bibnamefont {Behnke}},\ and\ \bibinfo {author} {\bibfnamefont
  {R.~J.}\ \bibnamefont {Thomas}},\ }\href
  {https://doi.org/10.1029/2012GL053666} {\bibfield  {journal} {\bibinfo
  {journal} {Geophysical Research Letters}\ }\textbf {\bibinfo {volume} {39}}
  (\bibinfo {year} {2012})}\BibitemShut {NoStop}%
\bibitem [{\citenamefont {Rhodes}\ \emph {et~al.}(1994)\citenamefont {Rhodes},
  \citenamefont {Shao}, \citenamefont {Krehbiel}, \citenamefont {Thomas},\ and\
  \citenamefont {Hayenga}}]{Rhodes:1994}%
  \BibitemOpen
  \bibfield  {author} {\bibinfo {author} {\bibfnamefont {C.~T.}\ \bibnamefont
  {Rhodes}}, \bibinfo {author} {\bibfnamefont {X.~M.}\ \bibnamefont {Shao}},
  \bibinfo {author} {\bibfnamefont {P.~R.}\ \bibnamefont {Krehbiel}}, \bibinfo
  {author} {\bibfnamefont {R.~J.}\ \bibnamefont {Thomas}},\ and\ \bibinfo
  {author} {\bibfnamefont {C.~O.}\ \bibnamefont {Hayenga}},\ }\href
  {https://doi.org/10.1029/94JD00318} {\bibfield  {journal} {\bibinfo
  {journal} {Journal of Geophysical Research: Atmospheres}\ }\textbf {\bibinfo
  {volume} {99}},\ \bibinfo {pages} {13059} (\bibinfo {year}
  {1994})}\BibitemShut {NoStop}%
\bibitem [{\citenamefont {Yoshida}\ \emph {et~al.}(2010)\citenamefont
  {Yoshida}, \citenamefont {Biagi}, \citenamefont {Rakov}, \citenamefont
  {Hill}, \citenamefont {Stapleton}, \citenamefont {Jordan}, \citenamefont
  {Uman}, \citenamefont {Morimoto}, \citenamefont {Ushio},\ and\ \citenamefont
  {Kawasaki}}]{Yoshida:2010}%
  \BibitemOpen
  \bibfield  {author} {\bibinfo {author} {\bibfnamefont {S.}~\bibnamefont
  {Yoshida}}, \bibinfo {author} {\bibfnamefont {C.~J.}\ \bibnamefont {Biagi}},
  \bibinfo {author} {\bibfnamefont {V.~A.}\ \bibnamefont {Rakov}}, \bibinfo
  {author} {\bibfnamefont {J.~D.}\ \bibnamefont {Hill}}, \bibinfo {author}
  {\bibfnamefont {M.~V.}\ \bibnamefont {Stapleton}}, \bibinfo {author}
  {\bibfnamefont {D.~M.}\ \bibnamefont {Jordan}}, \bibinfo {author}
  {\bibfnamefont {M.~A.}\ \bibnamefont {Uman}}, \bibinfo {author}
  {\bibfnamefont {T.}~\bibnamefont {Morimoto}}, \bibinfo {author}
  {\bibfnamefont {T.}~\bibnamefont {Ushio}},\ and\ \bibinfo {author}
  {\bibfnamefont {Z.-I.}\ \bibnamefont {Kawasaki}},\ }\href
  {https://doi.org/10.1029/2009GL042065} {\bibfield  {journal} {\bibinfo
  {journal} {Geophysical Research Letters}\ }\textbf {\bibinfo {volume} {37}}
  (\bibinfo {year} {2010})}\BibitemShut {NoStop}%
\bibitem [{\citenamefont {Stock}\ \emph {et~al.}(2014)\citenamefont {Stock},
  \citenamefont {Akita}, \citenamefont {Krehbiel}, \citenamefont {Rison},
  \citenamefont {Edens}, \citenamefont {Kawasaki},\ and\ \citenamefont
  {Stanley}}]{Stock:2014}%
  \BibitemOpen
  \bibfield  {author} {\bibinfo {author} {\bibfnamefont {M.~G.}\ \bibnamefont
  {Stock}}, \bibinfo {author} {\bibfnamefont {M.}~\bibnamefont {Akita}},
  \bibinfo {author} {\bibfnamefont {P.~R.}\ \bibnamefont {Krehbiel}}, \bibinfo
  {author} {\bibfnamefont {W.}~\bibnamefont {Rison}}, \bibinfo {author}
  {\bibfnamefont {H.~E.}\ \bibnamefont {Edens}}, \bibinfo {author}
  {\bibfnamefont {Z.}~\bibnamefont {Kawasaki}},\ and\ \bibinfo {author}
  {\bibfnamefont {M.~A.}\ \bibnamefont {Stanley}},\ }\href
  {https://doi.org/10.1002/2013JD020217} {\bibfield  {journal} {\bibinfo
  {journal} {Journal of Geophysical Research: Atmospheres}\ }\textbf {\bibinfo
  {volume} {119}},\ \bibinfo {pages} {3134} (\bibinfo {year}
  {2014})}\BibitemShut {NoStop}%
\bibitem [{\citenamefont {van Haarlem}\ \emph {et~al.}(2013)\citenamefont {van
  Haarlem} \emph {et~al.}}]{Haarlem:2013}%
  \BibitemOpen
  \bibfield  {author} {\bibinfo {author} {\bibfnamefont {M.~P.}\ \bibnamefont
  {van Haarlem}} \emph {et~al.},\ }\href
  {https://doi.org/10.1051/0004-6361/201220873} {\bibfield  {journal} {\bibinfo
   {journal} {A\&A}\ }\textbf {\bibinfo {volume} {556}},\ \bibinfo {pages} {A2}
  (\bibinfo {year} {2013})}\BibitemShut {NoStop}%
\bibitem [{\citenamefont {Hare}\ \emph {et~al.}(2019)\citenamefont {Hare},
  \citenamefont {Scholten}, \citenamefont {Dwyer}, \citenamefont {Trinh},
  \citenamefont {Buitink}, \citenamefont {ter Veen}, \citenamefont {Bonardi},
  \citenamefont {Corstanje}, \citenamefont {Falcke}, \citenamefont
  {H\"{o}randel}, \citenamefont {Huege}, \citenamefont {Mitra}, \citenamefont
  {Mulrey}, \citenamefont {Nelles}, \citenamefont {Rachen}, \citenamefont
  {Rossetto}, \citenamefont {Schellart}, \citenamefont {Winchen}, \citenamefont
  {Anderson}, \citenamefont {Avruch}, \citenamefont {Bentum}, \citenamefont
  {Blaauw}, \citenamefont {Broderick}, \citenamefont {Brouw}, \citenamefont
  {Br\"{u}ggen}, \citenamefont {Butcher}, \citenamefont {Ciardi}, \citenamefont
  {Fallows}, \citenamefont {de~Geus}, \citenamefont {Duscha}, \citenamefont
  {Eisl\"{o}ffel}, \citenamefont {Garrett}, \citenamefont {Grie{\ss}meier},
  \citenamefont {Gunst}, \citenamefont {van Haarlem}, \citenamefont {Hessels},
  \citenamefont {Hoeft}, \citenamefont {van~der Horst}, \citenamefont
  {Iacobelli}, \citenamefont {Koopmans}, \citenamefont {Krankowski},
  \citenamefont {Maat}, \citenamefont {Norden}, \citenamefont {Paas},
  \citenamefont {Pandey-Pommier}, \citenamefont {Pandey}, \citenamefont
  {Pekal}, \citenamefont {Pizzo}, \citenamefont {Reich}, \citenamefont
  {Rothkaehl}, \citenamefont {R\"{o}ttgering}, \citenamefont {Rowlinson},
  \citenamefont {Schwarz}, \citenamefont {Shulevski}, \citenamefont {Sluman},
  \citenamefont {Smirnov}, \citenamefont {Soida}, \citenamefont {Tagger},
  \citenamefont {Toribio}, \citenamefont {van Ardenne}, \citenamefont {Wijers},
  \citenamefont {van Weeren}, \citenamefont {Wucknitz}, \citenamefont {Zarka},\
  and\ \citenamefont {Zucca}}]{Hare:2019}%
  \BibitemOpen
  \bibfield  {author} {\bibinfo {author} {\bibfnamefont {B.~M.}\ \bibnamefont
  {Hare}}, \bibinfo {author} {\bibfnamefont {O.}~\bibnamefont {Scholten}},
  \bibinfo {author} {\bibfnamefont {J.}~\bibnamefont {Dwyer}}, \bibinfo
  {author} {\bibfnamefont {T.~N.~G.}\ \bibnamefont {Trinh}}, \bibinfo {author}
  {\bibfnamefont {S.}~\bibnamefont {Buitink}}, \bibinfo {author} {\bibfnamefont
  {S.}~\bibnamefont {ter Veen}}, \bibinfo {author} {\bibfnamefont
  {A.}~\bibnamefont {Bonardi}}, \bibinfo {author} {\bibfnamefont
  {A.}~\bibnamefont {Corstanje}}, \bibinfo {author} {\bibfnamefont
  {H.}~\bibnamefont {Falcke}}, \bibinfo {author} {\bibfnamefont {J.~R.}\
  \bibnamefont {H\"{o}randel}}, \bibinfo {author} {\bibfnamefont
  {T.}~\bibnamefont {Huege}}, \bibinfo {author} {\bibfnamefont
  {P.}~\bibnamefont {Mitra}}, \bibinfo {author} {\bibfnamefont
  {K.}~\bibnamefont {Mulrey}}, \bibinfo {author} {\bibfnamefont
  {A.}~\bibnamefont {Nelles}}, \bibinfo {author} {\bibfnamefont {J.~P.}\
  \bibnamefont {Rachen}}, \bibinfo {author} {\bibfnamefont {L.}~\bibnamefont
  {Rossetto}}, \bibinfo {author} {\bibfnamefont {P.}~\bibnamefont {Schellart}},
  \bibinfo {author} {\bibfnamefont {T.}~\bibnamefont {Winchen}}, \bibinfo
  {author} {\bibfnamefont {J.}~\bibnamefont {Anderson}}, \bibinfo {author}
  {\bibfnamefont {I.~M.}\ \bibnamefont {Avruch}}, \bibinfo {author}
  {\bibfnamefont {M.~J.}\ \bibnamefont {Bentum}}, \bibinfo {author}
  {\bibfnamefont {R.}~\bibnamefont {Blaauw}}, \bibinfo {author} {\bibfnamefont
  {J.~W.}\ \bibnamefont {Broderick}}, \bibinfo {author} {\bibfnamefont {W.~N.}\
  \bibnamefont {Brouw}}, \bibinfo {author} {\bibfnamefont {M.}~\bibnamefont
  {Br\"{u}ggen}}, \bibinfo {author} {\bibfnamefont {H.~R.}\ \bibnamefont
  {Butcher}}, \bibinfo {author} {\bibfnamefont {B.}~\bibnamefont {Ciardi}},
  \bibinfo {author} {\bibfnamefont {R.~A.}\ \bibnamefont {Fallows}}, \bibinfo
  {author} {\bibfnamefont {E.}~\bibnamefont {de~Geus}}, \bibinfo {author}
  {\bibfnamefont {S.}~\bibnamefont {Duscha}}, \bibinfo {author} {\bibfnamefont
  {J.}~\bibnamefont {Eisl\"{o}ffel}}, \bibinfo {author} {\bibfnamefont {M.~A.}\
  \bibnamefont {Garrett}}, \bibinfo {author} {\bibfnamefont {J.~M.}\
  \bibnamefont {Grie{\ss}meier}}, \bibinfo {author} {\bibfnamefont {A.~W.}\
  \bibnamefont {Gunst}}, \bibinfo {author} {\bibfnamefont {M.~P.}\ \bibnamefont
  {van Haarlem}}, \bibinfo {author} {\bibfnamefont {J.~W.~T.}\ \bibnamefont
  {Hessels}}, \bibinfo {author} {\bibfnamefont {M.}~\bibnamefont {Hoeft}},
  \bibinfo {author} {\bibfnamefont {A.~J.}\ \bibnamefont {van~der Horst}},
  \bibinfo {author} {\bibfnamefont {M.}~\bibnamefont {Iacobelli}}, \bibinfo
  {author} {\bibfnamefont {L.~V.~E.}\ \bibnamefont {Koopmans}}, \bibinfo
  {author} {\bibfnamefont {A.}~\bibnamefont {Krankowski}}, \bibinfo {author}
  {\bibfnamefont {P.}~\bibnamefont {Maat}}, \bibinfo {author} {\bibfnamefont
  {M.~J.}\ \bibnamefont {Norden}}, \bibinfo {author} {\bibfnamefont
  {H.}~\bibnamefont {Paas}}, \bibinfo {author} {\bibfnamefont {M.}~\bibnamefont
  {Pandey-Pommier}}, \bibinfo {author} {\bibfnamefont {V.~N.}\ \bibnamefont
  {Pandey}}, \bibinfo {author} {\bibfnamefont {R.}~\bibnamefont {Pekal}},
  \bibinfo {author} {\bibfnamefont {R.}~\bibnamefont {Pizzo}}, \bibinfo
  {author} {\bibfnamefont {W.}~\bibnamefont {Reich}}, \bibinfo {author}
  {\bibfnamefont {H.}~\bibnamefont {Rothkaehl}}, \bibinfo {author}
  {\bibfnamefont {H.~J.~A.}\ \bibnamefont {R\"{o}ttgering}}, \bibinfo {author}
  {\bibfnamefont {A.}~\bibnamefont {Rowlinson}}, \bibinfo {author}
  {\bibfnamefont {D.~J.}\ \bibnamefont {Schwarz}}, \bibinfo {author}
  {\bibfnamefont {A.}~\bibnamefont {Shulevski}}, \bibinfo {author}
  {\bibfnamefont {J.}~\bibnamefont {Sluman}}, \bibinfo {author} {\bibfnamefont
  {O.}~\bibnamefont {Smirnov}}, \bibinfo {author} {\bibfnamefont
  {M.}~\bibnamefont {Soida}}, \bibinfo {author} {\bibfnamefont
  {M.}~\bibnamefont {Tagger}}, \bibinfo {author} {\bibfnamefont {M.~C.}\
  \bibnamefont {Toribio}}, \bibinfo {author} {\bibfnamefont {A.}~\bibnamefont
  {van Ardenne}}, \bibinfo {author} {\bibfnamefont {R.~A. M.~J.}\ \bibnamefont
  {Wijers}}, \bibinfo {author} {\bibfnamefont {R.~J.}\ \bibnamefont {van
  Weeren}}, \bibinfo {author} {\bibfnamefont {O.}~\bibnamefont {Wucknitz}},
  \bibinfo {author} {\bibfnamefont {P.}~\bibnamefont {Zarka}},\ and\ \bibinfo
  {author} {\bibfnamefont {P.}~\bibnamefont {Zucca}},\ }\href
  {https://doi.org/10.1038/s41586-019-1086-6} {\bibfield  {journal} {\bibinfo
  {journal} {Nature}\ }\textbf {\bibinfo {volume} {568}},\ \bibinfo {pages}
  {360} (\bibinfo {year} {2019})}\BibitemShut {NoStop}%
\bibitem [{\citenamefont {Scholten}\ \emph
  {et~al.}(2021{\natexlab{a}})\citenamefont {Scholten}, \citenamefont {Hare},
  \citenamefont {Dwyer}, \citenamefont {Sterpka}, \citenamefont {Kolmasova},
  \citenamefont {Santolik}, \citenamefont {Lan}, \citenamefont {Uhlir},
  \citenamefont {Buitink}, \citenamefont {Corstanje}, \citenamefont {Falcke},
  \citenamefont {Huege}, \citenamefont {Hoerandel}, \citenamefont {Krampah},
  \citenamefont {Mitra}, \citenamefont {Mulrey}, \citenamefont {Nelles},
  \citenamefont {Pandya}, \citenamefont {Pel}, \citenamefont {Rachen},
  \citenamefont {Trinh}, \citenamefont {Veen}, \citenamefont {Thoudam},\ and\
  \citenamefont {Winchen}}]{Scholten:2021-init}%
  \BibitemOpen
  \bibfield  {author} {\bibinfo {author} {\bibfnamefont {O.}~\bibnamefont
  {Scholten}}, \bibinfo {author} {\bibfnamefont {B.~M.}\ \bibnamefont {Hare}},
  \bibinfo {author} {\bibfnamefont {J.}~\bibnamefont {Dwyer}}, \bibinfo
  {author} {\bibfnamefont {C.}~\bibnamefont {Sterpka}}, \bibinfo {author}
  {\bibfnamefont {I.}~\bibnamefont {Kolmasova}}, \bibinfo {author}
  {\bibfnamefont {O.}~\bibnamefont {Santolik}}, \bibinfo {author}
  {\bibfnamefont {R.}~\bibnamefont {Lan}}, \bibinfo {author} {\bibfnamefont
  {L.}~\bibnamefont {Uhlir}}, \bibinfo {author} {\bibfnamefont
  {S.}~\bibnamefont {Buitink}}, \bibinfo {author} {\bibfnamefont
  {A.}~\bibnamefont {Corstanje}}, \bibinfo {author} {\bibfnamefont
  {H.}~\bibnamefont {Falcke}}, \bibinfo {author} {\bibfnamefont
  {T.}~\bibnamefont {Huege}}, \bibinfo {author} {\bibfnamefont {J.~R.}\
  \bibnamefont {Hoerandel}}, \bibinfo {author} {\bibfnamefont {G.~K.}\
  \bibnamefont {Krampah}}, \bibinfo {author} {\bibfnamefont {P.}~\bibnamefont
  {Mitra}}, \bibinfo {author} {\bibfnamefont {K.}~\bibnamefont {Mulrey}},
  \bibinfo {author} {\bibfnamefont {A.}~\bibnamefont {Nelles}}, \bibinfo
  {author} {\bibfnamefont {H.}~\bibnamefont {Pandya}}, \bibinfo {author}
  {\bibfnamefont {A.}~\bibnamefont {Pel}}, \bibinfo {author} {\bibfnamefont
  {J.~P.}\ \bibnamefont {Rachen}}, \bibinfo {author} {\bibfnamefont {T.~N.~G.}\
  \bibnamefont {Trinh}}, \bibinfo {author} {\bibfnamefont {S.~t.}\ \bibnamefont
  {Veen}}, \bibinfo {author} {\bibfnamefont {S.}~\bibnamefont {Thoudam}},\ and\
  \bibinfo {author} {\bibfnamefont {T.}~\bibnamefont {Winchen}},\ }\href
  {https://doi.org/https://doi.org/10.1029/2020JD033126} {\bibfield  {journal}
  {\bibinfo  {journal} {Journal of Geophysical Research: Atmospheres}\ }\textbf
  {\bibinfo {volume} {126}},\ \bibinfo {pages} {e2020JD033126} (\bibinfo {year}
  {2021}{\natexlab{a}})},\ \bibinfo {note} {e2020JD033126
  2020JD033126}\BibitemShut {NoStop}%
\bibitem [{\citenamefont {Scholten}\ \emph
  {et~al.}(2021{\natexlab{b}})\citenamefont {Scholten}, \citenamefont {Hare},
  \citenamefont {Dwyer}, \citenamefont {Liu}, \citenamefont {Sterpka},
  \citenamefont {Kolmasova}, \citenamefont {Santolik}, \citenamefont {Lan},
  \citenamefont {Uhlir}, \citenamefont {Buitink}, \citenamefont {Corstanje},
  \citenamefont {Falcke}, \citenamefont {Huege}, \citenamefont {Hoerandel},
  \citenamefont {Krampah}, \citenamefont {Mitra}, \citenamefont {Mulrey},
  \citenamefont {Nelles}, \citenamefont {Pandya}, \citenamefont {Rachen},
  \citenamefont {Trinh}, \citenamefont {Veen}, \citenamefont {Thoudam},\ and\
  \citenamefont {Winchen}}]{Scholten:2021-RNL}%
  \BibitemOpen
  \bibfield  {author} {\bibinfo {author} {\bibfnamefont {O.}~\bibnamefont
  {Scholten}}, \bibinfo {author} {\bibfnamefont {B.~M.}\ \bibnamefont {Hare}},
  \bibinfo {author} {\bibfnamefont {J.}~\bibnamefont {Dwyer}}, \bibinfo
  {author} {\bibfnamefont {N.}~\bibnamefont {Liu}}, \bibinfo {author}
  {\bibfnamefont {C.}~\bibnamefont {Sterpka}}, \bibinfo {author} {\bibfnamefont
  {I.}~\bibnamefont {Kolmasova}}, \bibinfo {author} {\bibfnamefont
  {O.}~\bibnamefont {Santolik}}, \bibinfo {author} {\bibfnamefont
  {R.}~\bibnamefont {Lan}}, \bibinfo {author} {\bibfnamefont {L.}~\bibnamefont
  {Uhlir}}, \bibinfo {author} {\bibfnamefont {S.}~\bibnamefont {Buitink}},
  \bibinfo {author} {\bibfnamefont {A.}~\bibnamefont {Corstanje}}, \bibinfo
  {author} {\bibfnamefont {H.}~\bibnamefont {Falcke}}, \bibinfo {author}
  {\bibfnamefont {T.}~\bibnamefont {Huege}}, \bibinfo {author} {\bibfnamefont
  {J.~R.}\ \bibnamefont {Hoerandel}}, \bibinfo {author} {\bibfnamefont {G.~K.}\
  \bibnamefont {Krampah}}, \bibinfo {author} {\bibfnamefont {P.}~\bibnamefont
  {Mitra}}, \bibinfo {author} {\bibfnamefont {K.}~\bibnamefont {Mulrey}},
  \bibinfo {author} {\bibfnamefont {A.}~\bibnamefont {Nelles}}, \bibinfo
  {author} {\bibfnamefont {H.}~\bibnamefont {Pandya}}, \bibinfo {author}
  {\bibfnamefont {J.~P.}\ \bibnamefont {Rachen}}, \bibinfo {author}
  {\bibfnamefont {T.~N.~G.}\ \bibnamefont {Trinh}}, \bibinfo {author}
  {\bibfnamefont {S.~t.}\ \bibnamefont {Veen}}, \bibinfo {author}
  {\bibfnamefont {S.}~\bibnamefont {Thoudam}},\ and\ \bibinfo {author}
  {\bibfnamefont {T.}~\bibnamefont {Winchen}},\ }\href
  {https://doi.org/10.1038/s41598-021-95433-5} {\bibfield  {journal} {\bibinfo
  {journal} {Scientific Reports}\ }\textbf {\bibinfo {volume} {11}},\ \bibinfo
  {pages} {16256} (\bibinfo {year} {2021}{\natexlab{b}})}\BibitemShut {NoStop}%
\bibitem [{\citenamefont {Scholten}\ \emph
  {et~al.}(2021{\natexlab{c}})\citenamefont {Scholten}, \citenamefont {Hare},
  \citenamefont {Dwyer}, \citenamefont {Liu}, \citenamefont {Sterpka},
  \citenamefont {Buitink}, \citenamefont {Huege}, \citenamefont {Nelles},\ and\
  \citenamefont {ter Veen}}]{Scholten:2021-INL}%
  \BibitemOpen
  \bibfield  {author} {\bibinfo {author} {\bibfnamefont {O.}~\bibnamefont
  {Scholten}}, \bibinfo {author} {\bibfnamefont {B.~M.}\ \bibnamefont {Hare}},
  \bibinfo {author} {\bibfnamefont {J.}~\bibnamefont {Dwyer}}, \bibinfo
  {author} {\bibfnamefont {N.}~\bibnamefont {Liu}}, \bibinfo {author}
  {\bibfnamefont {C.}~\bibnamefont {Sterpka}}, \bibinfo {author} {\bibfnamefont
  {S.}~\bibnamefont {Buitink}}, \bibinfo {author} {\bibfnamefont
  {T.}~\bibnamefont {Huege}}, \bibinfo {author} {\bibfnamefont
  {A.}~\bibnamefont {Nelles}},\ and\ \bibinfo {author} {\bibfnamefont
  {S.}~\bibnamefont {ter Veen}},\ }\href
  {https://doi.org/10.1103/PhysRevD.104.063022} {\bibfield  {journal} {\bibinfo
   {journal} {Phys. Rev. D}\ }\textbf {\bibinfo {volume} {104}},\ \bibinfo
  {pages} {063022} (\bibinfo {year} {2021}{\natexlab{c}})}\BibitemShut
  {NoStop}%
\bibitem [{\citenamefont {Hare}\ \emph {et~al.}(2020)\citenamefont {Hare},
  \citenamefont {Scholten}, \citenamefont {Dwyer}, \citenamefont {Ebert},
  \citenamefont {Nijdam}, \citenamefont {Bonardi}, \citenamefont {Buitink},
  \citenamefont {Corstanje}, \citenamefont {Falcke}, \citenamefont {Huege},
  \citenamefont {H\"orandel}, \citenamefont {Krampah}, \citenamefont {Mitra},
  \citenamefont {Mulrey}, \citenamefont {Neijzen}, \citenamefont {Nelles},
  \citenamefont {Pandya}, \citenamefont {Rachen}, \citenamefont {Rossetto},
  \citenamefont {Trinh}, \citenamefont {ter Veen},\ and\ \citenamefont
  {Winchen}}]{Hare:2020}%
  \BibitemOpen
  \bibfield  {author} {\bibinfo {author} {\bibfnamefont {B.~M.}\ \bibnamefont
  {Hare}}, \bibinfo {author} {\bibfnamefont {O.}~\bibnamefont {Scholten}},
  \bibinfo {author} {\bibfnamefont {J.}~\bibnamefont {Dwyer}}, \bibinfo
  {author} {\bibfnamefont {U.}~\bibnamefont {Ebert}}, \bibinfo {author}
  {\bibfnamefont {S.}~\bibnamefont {Nijdam}}, \bibinfo {author} {\bibfnamefont
  {A.}~\bibnamefont {Bonardi}}, \bibinfo {author} {\bibfnamefont
  {S.}~\bibnamefont {Buitink}}, \bibinfo {author} {\bibfnamefont
  {A.}~\bibnamefont {Corstanje}}, \bibinfo {author} {\bibfnamefont
  {H.}~\bibnamefont {Falcke}}, \bibinfo {author} {\bibfnamefont
  {T.}~\bibnamefont {Huege}}, \bibinfo {author} {\bibfnamefont {J.~R.}\
  \bibnamefont {H\"orandel}}, \bibinfo {author} {\bibfnamefont {G.~K.}\
  \bibnamefont {Krampah}}, \bibinfo {author} {\bibfnamefont {P.}~\bibnamefont
  {Mitra}}, \bibinfo {author} {\bibfnamefont {K.}~\bibnamefont {Mulrey}},
  \bibinfo {author} {\bibfnamefont {B.}~\bibnamefont {Neijzen}}, \bibinfo
  {author} {\bibfnamefont {A.}~\bibnamefont {Nelles}}, \bibinfo {author}
  {\bibfnamefont {H.}~\bibnamefont {Pandya}}, \bibinfo {author} {\bibfnamefont
  {J.~P.}\ \bibnamefont {Rachen}}, \bibinfo {author} {\bibfnamefont
  {L.}~\bibnamefont {Rossetto}}, \bibinfo {author} {\bibfnamefont {T.~N.~G.}\
  \bibnamefont {Trinh}}, \bibinfo {author} {\bibfnamefont {S.}~\bibnamefont
  {ter Veen}},\ and\ \bibinfo {author} {\bibfnamefont {T.}~\bibnamefont
  {Winchen}},\ }\href {https://doi.org/10.1103/PhysRevLett.124.105101}
  {\bibfield  {journal} {\bibinfo  {journal} {Phys. Rev. Lett.}\ }\textbf
  {\bibinfo {volume} {124}},\ \bibinfo {pages} {105101} (\bibinfo {year}
  {2020})}\BibitemShut {NoStop}%
\bibitem [{\citenamefont {Gurevich}\ \emph {et~al.}(1992)\citenamefont
  {Gurevich}, \citenamefont {Milikh},\ and\ \citenamefont
  {Roussel-Dupre}}]{Gurevich:1992}%
  \BibitemOpen
  \bibfield  {author} {\bibinfo {author} {\bibfnamefont {A.}~\bibnamefont
  {Gurevich}}, \bibinfo {author} {\bibfnamefont {G.}~\bibnamefont {Milikh}},\
  and\ \bibinfo {author} {\bibfnamefont {R.}~\bibnamefont {Roussel-Dupre}},\
  }\href {https://doi.org/10.1016/0375-9601(92)90348-P} {\bibfield  {journal}
  {\bibinfo  {journal} {Physics Letters A}\ }\textbf {\bibinfo {volume}
  {165}},\ \bibinfo {pages} {463 } (\bibinfo {year} {1992})}\BibitemShut
  {NoStop}%
\bibitem [{\citenamefont {Enoto}\ \emph {et~al.}(2017)\citenamefont {Enoto}
  \emph {et~al.}}]{Enoto:2017}%
  \BibitemOpen
  \bibfield  {author} {\bibinfo {author} {\bibfnamefont {T.}~\bibnamefont
  {Enoto}} \emph {et~al.},\ }\href {https://doi.org/10.1038/nature24630}
  {\bibfield  {journal} {\bibinfo  {journal} {Nature}\ }\textbf {\bibinfo
  {volume} {551}},\ \bibinfo {pages} {481} (\bibinfo {year}
  {2017})}\BibitemShut {NoStop}%
\bibitem [{\citenamefont {Neubert}\ \emph {et~al.}(2020)\citenamefont
  {Neubert}, \citenamefont {{\O}stgaard}, \citenamefont {Reglero},
  \citenamefont {Chanrion}, \citenamefont {Heumesser}, \citenamefont
  {Dimitriadou}, \citenamefont {Christiansen}, \citenamefont
  {Budtz-J{\o}rgensen}, \citenamefont {Kuvvetli}, \citenamefont {Rasmussen},
  \citenamefont {Mezentsev}, \citenamefont {Marisaldi}, \citenamefont
  {Ullaland}, \citenamefont {Genov}, \citenamefont {Yang}, \citenamefont
  {Kochkin}, \citenamefont {Navarro-Gonzalez}, \citenamefont {Connell},\ and\
  \citenamefont {Eyles}}]{Neubert:2020}%
  \BibitemOpen
  \bibfield  {author} {\bibinfo {author} {\bibfnamefont {T.}~\bibnamefont
  {Neubert}}, \bibinfo {author} {\bibfnamefont {N.}~\bibnamefont
  {{\O}stgaard}}, \bibinfo {author} {\bibfnamefont {V.}~\bibnamefont
  {Reglero}}, \bibinfo {author} {\bibfnamefont {O.}~\bibnamefont {Chanrion}},
  \bibinfo {author} {\bibfnamefont {M.}~\bibnamefont {Heumesser}}, \bibinfo
  {author} {\bibfnamefont {K.}~\bibnamefont {Dimitriadou}}, \bibinfo {author}
  {\bibfnamefont {F.}~\bibnamefont {Christiansen}}, \bibinfo {author}
  {\bibfnamefont {C.}~\bibnamefont {Budtz-J{\o}rgensen}}, \bibinfo {author}
  {\bibfnamefont {I.}~\bibnamefont {Kuvvetli}}, \bibinfo {author}
  {\bibfnamefont {I.~L.}\ \bibnamefont {Rasmussen}}, \bibinfo {author}
  {\bibfnamefont {A.}~\bibnamefont {Mezentsev}}, \bibinfo {author}
  {\bibfnamefont {M.}~\bibnamefont {Marisaldi}}, \bibinfo {author}
  {\bibfnamefont {K.}~\bibnamefont {Ullaland}}, \bibinfo {author}
  {\bibfnamefont {G.}~\bibnamefont {Genov}}, \bibinfo {author} {\bibfnamefont
  {S.}~\bibnamefont {Yang}}, \bibinfo {author} {\bibfnamefont {P.}~\bibnamefont
  {Kochkin}}, \bibinfo {author} {\bibfnamefont {J.}~\bibnamefont
  {Navarro-Gonzalez}}, \bibinfo {author} {\bibfnamefont {P.~H.}\ \bibnamefont
  {Connell}},\ and\ \bibinfo {author} {\bibfnamefont {C.~J.}\ \bibnamefont
  {Eyles}},\ }\href {https://doi.org/10.1126/science.aax3872} {\bibfield
  {journal} {\bibinfo  {journal} {Science}\ }\textbf {\bibinfo {volume}
  {367}},\ \bibinfo {pages} {183} (\bibinfo {year} {2020})},\ \Eprint
  {https://arxiv.org/abs/https://science.sciencemag.org/content/367/6474/183.full.pdf}
  {https://science.sciencemag.org/content/367/6474/183.full.pdf} \BibitemShut
  {NoStop}%
\bibitem [{\citenamefont {Gibney}(2021)}]{Gibney:2021}%
  \BibitemOpen
  \bibfield  {author} {\bibinfo {author} {\bibfnamefont {E.}~\bibnamefont
  {Gibney}},\ }\href {https://doi.org/10.1038/d41586-021-00395-3} {\bibfield
  {journal} {\bibinfo  {journal} {Nature}\ }\textbf {\bibinfo {volume} {590}},\
  \bibinfo {pages} {378} (\bibinfo {year} {2021})}\BibitemShut {NoStop}%
\bibitem [{\citenamefont {Maiorana}\ \emph {et~al.}()\citenamefont {Maiorana},
  \citenamefont {Marisaldi}, \citenamefont {F\"{u}llekrug}, \citenamefont
  {Soula}, \citenamefont {Lapierre}, \citenamefont {Mezentsev}, \citenamefont
  {Skeie}, \citenamefont {Heumesser}, \citenamefont {Chanrion}, \citenamefont
  {{\O}stgaard}, \citenamefont {Neubert},\ and\ \citenamefont
  {Reglero}}]{Maiorana:2021}%
  \BibitemOpen
  \bibfield  {author} {\bibinfo {author} {\bibfnamefont {C.}~\bibnamefont
  {Maiorana}}, \bibinfo {author} {\bibfnamefont {M.}~\bibnamefont {Marisaldi}},
  \bibinfo {author} {\bibfnamefont {M.}~\bibnamefont {F\"{u}llekrug}}, \bibinfo
  {author} {\bibfnamefont {S.}~\bibnamefont {Soula}}, \bibinfo {author}
  {\bibfnamefont {J.}~\bibnamefont {Lapierre}}, \bibinfo {author}
  {\bibfnamefont {A.}~\bibnamefont {Mezentsev}}, \bibinfo {author}
  {\bibfnamefont {C.~A.}\ \bibnamefont {Skeie}}, \bibinfo {author}
  {\bibfnamefont {M.}~\bibnamefont {Heumesser}}, \bibinfo {author}
  {\bibfnamefont {O.}~\bibnamefont {Chanrion}}, \bibinfo {author}
  {\bibfnamefont {N.}~\bibnamefont {{\O}stgaard}}, \bibinfo {author}
  {\bibfnamefont {T.}~\bibnamefont {Neubert}},\ and\ \bibinfo {author}
  {\bibfnamefont {V.}~\bibnamefont {Reglero}},\ }\href
  {https://doi.org/https://doi.org/10.1029/2020JD034432} {\bibfield  {journal}
  {\bibinfo  {journal} {Journal of Geophysical Research: Atmospheres}\ }\textbf
  {\bibinfo {volume} {n/a}},\ \bibinfo {pages} {e2020JD034432}}\BibitemShut
  {NoStop}%
\bibitem [{\citenamefont {Jackson}(1975)}]{Jackson:1975}%
  \BibitemOpen
  \bibfield  {author} {\bibinfo {author} {\bibfnamefont {J.~D.}\ \bibnamefont
  {Jackson}},\ }\href {https://cds.cern.ch/record/100964} {\emph {\bibinfo
  {title} {{Classical electrodynamics; 2nd ed.}}}}\ (\bibinfo  {publisher}
  {Wiley},\ \bibinfo {address} {New York, NY},\ \bibinfo {year}
  {1975})\BibitemShut {NoStop}%
\bibitem [{\citenamefont {Ka\v{s}par}\ \emph {et~al.}(2015)\citenamefont
  {Ka\v{s}par}, \citenamefont {Santol\'{i}k},\ and\ \citenamefont
  {Kolma\v{s}ov\'{a}}}]{Kaspar:2015}%
  \BibitemOpen
  \bibfield  {author} {\bibinfo {author} {\bibfnamefont {P.}~\bibnamefont
  {Ka\v{s}par}}, \bibinfo {author} {\bibfnamefont {O.}~\bibnamefont
  {Santol\'{i}k}},\ and\ \bibinfo {author} {\bibfnamefont {I.}~\bibnamefont
  {Kolma\v{s}ov\'{a}}},\ }\href {https://doi.org/10.1002/2015GL064777}
  {\bibfield  {journal} {\bibinfo  {journal} {Geophysical Research Letters}\
  }\textbf {\bibinfo {volume} {42}},\ \bibinfo {pages} {7206} (\bibinfo {year}
  {2015})},\ \Eprint
  {https://arxiv.org/abs/https://agupubs.onlinelibrary.wiley.com/doi/pdf/10.1002/2015GL064777}
  {https://agupubs.onlinelibrary.wiley.com/doi/pdf/10.1002/2015GL064777}
  \BibitemShut {NoStop}%
\bibitem [{\citenamefont {Uman}(2001)}]{Uman:2001}%
  \BibitemOpen
  \bibfield  {author} {\bibinfo {author} {\bibfnamefont {M.~A.}\ \bibnamefont
  {Uman}},\ }\href@noop {} {\emph {\bibinfo {title} {The Lightning
  Discharge}}}\ (\bibinfo  {publisher} {Dover Publications, Mineola, New
  York},\ \bibinfo {year} {2001})\BibitemShut {NoStop}%
\bibitem [{\citenamefont {Wada}\ \emph {et~al.}(2019)\citenamefont {Wada},
  \citenamefont {Enoto}, \citenamefont {Nakazawa}, \citenamefont {Furuta},
  \citenamefont {Yuasa}, \citenamefont {Nakamura}, \citenamefont {Morimoto},
  \citenamefont {Matsumoto}, \citenamefont {Makishima},\ and\ \citenamefont
  {Tsuchiya}}]{Wada:2019}%
  \BibitemOpen
  \bibfield  {author} {\bibinfo {author} {\bibfnamefont {Y.}~\bibnamefont
  {Wada}}, \bibinfo {author} {\bibfnamefont {T.}~\bibnamefont {Enoto}},
  \bibinfo {author} {\bibfnamefont {K.}~\bibnamefont {Nakazawa}}, \bibinfo
  {author} {\bibfnamefont {Y.}~\bibnamefont {Furuta}}, \bibinfo {author}
  {\bibfnamefont {T.}~\bibnamefont {Yuasa}}, \bibinfo {author} {\bibfnamefont
  {Y.}~\bibnamefont {Nakamura}}, \bibinfo {author} {\bibfnamefont
  {T.}~\bibnamefont {Morimoto}}, \bibinfo {author} {\bibfnamefont
  {T.}~\bibnamefont {Matsumoto}}, \bibinfo {author} {\bibfnamefont
  {K.}~\bibnamefont {Makishima}},\ and\ \bibinfo {author} {\bibfnamefont
  {H.}~\bibnamefont {Tsuchiya}},\ }\href
  {https://doi.org/10.1103/PhysRevLett.123.061103} {\bibfield  {journal}
  {\bibinfo  {journal} {Phys. Rev. Lett.}\ }\textbf {\bibinfo {volume} {123}},\
  \bibinfo {pages} {061103} (\bibinfo {year} {2019})}\BibitemShut {NoStop}%
\bibitem [{\citenamefont {ASTRON}(2020)}]{LOFAR-LTA}%
  \BibitemOpen
  \bibfield  {author} {\bibinfo {author} {\bibnamefont {ASTRON}},\ }\href@noop
  {} {\bibinfo {title} {{LOFAR Long Term Archive Access}}},\ \bibinfo
  {howpublished}
  {\url{https://www.astron.nl/lofarwiki/doku.php?id=public:lta_howto}}
  (\bibinfo {year} {2020})\BibitemShut {NoStop}%
\bibitem [{\citenamefont {Scholten}(2021)}]{Scholten-v20:2021}%
  \BibitemOpen
  \bibfield  {author} {\bibinfo {author} {\bibfnamefont {O.}~\bibnamefont
  {Scholten}},\ }\href {https://doi.org/10.5281/zenodo.5550959} {\bibinfo
  {title} {Lofar lightning imaging, v20}} (\bibinfo {year} {2021})\BibitemShut
  {NoStop}%
\bibitem [{\citenamefont {Pugmire}\ \emph {et~al.}(2015)\citenamefont
  {Pugmire}, \citenamefont {Mundt}, \citenamefont {LaBella},\ and\
  \citenamefont {Struyf}}]{GLE}%
  \BibitemOpen
  \bibfield  {author} {\bibinfo {author} {\bibfnamefont {C.}~\bibnamefont
  {Pugmire}}, \bibinfo {author} {\bibfnamefont {S.~M.}\ \bibnamefont {Mundt}},
  \bibinfo {author} {\bibfnamefont {V.~P.}\ \bibnamefont {LaBella}},\ and\
  \bibinfo {author} {\bibfnamefont {J.}~\bibnamefont {Struyf}},\ }\href
  {https://glx.sourceforge.io/index.html} {\bibinfo {title} {Graphics layout
  engine gle 4.2.5 user manual}},\ \bibinfo {howpublished}
  {\url{https://en.wikipedia.org/wiki/Graphics_Layout_Engine}} (\bibinfo {year}
  {2015}),\ \Eprint {https://arxiv.org/abs/http://www.gle-graphics.org/}
  {http://www.gle-graphics.org/} \BibitemShut {NoStop}%
\bibitem [{\citenamefont {Arts}\ and\ \citenamefont
  {Prasad}(2021)}]{Arts:2021}%
  \BibitemOpen
  \bibfield  {author} {\bibinfo {author} {\bibfnamefont {M.}~\bibnamefont
  {Arts}}\ and\ \bibinfo {author} {\bibfnamefont {P.}~\bibnamefont {Prasad}},\
  }\href {https://doi.org/10.1002/essoar.10507422.1} {\bibfield  {journal}
  {\bibinfo  {journal} {Earth and Space Science Open Archive}\ ,\ \bibinfo
  {pages} {30}} (\bibinfo {year} {2021})}\BibitemShut {NoStop}%
\end{thebibliography}%

\end{document}